%% file: Express2018.tex
\documentclass[submission,copyright,creativecommons]{eptcs}
\providecommand{\event}{EXPRESS/SOS 2018} % Name of the event you are submitting to
\usepackage{breakurl}             % Not needed if you use pdflatex only.
\usepackage{underscore}           % Only needed if you use pdflatex.
\usepackage{xcolor}
\usepackage{graphicx}
\usepackage[justification=centering]{caption}
\usepackage{listings}
\usepackage{float}
 
 \usepackage{enumitem}   

\usepackage{fontenc}
\usepackage{amsmath}
\usepackage{amsthm}
 
\newtheorem{definition}{Definition}
\newtheorem{theorem}{Theorem}

\title{A Classification of BPMN Collaborations based on\\ Safeness and Soundness Notions}
\author{Flavio Corradini, Chiara Muzi, Barbara Re, Francesco Tiezzi
\institute{Computer Science Division, School of Science and Technology\\
University of Camerino, Camerino, Italy}
\email{\{\emph{name.surname}\}@unicam.it}
}
\def\titlerunning{A Classification of BPMN Collaborations based on Safeness and Soundness Notions}
\def\authorrunning{F. Corradini, C. Muzi, B. Re \& F. Tiezzi}

\input{MacroClassification}

\input{annotationnote}
\input{macroProoftree}

\begin{document}

\maketitle

 \begin{abstract}
BPMN 2.0 standard has a huge uptake in modelling business processes within the same organisation or collaborations involving multiple interacting participants. It results that providing a solid foundation to enable BPMN designers to understand their models in a consistent way is becoming more and more important.  In our investigation we define and exploit a formal characterisation of the collaborations' semantics, specifically and directly given for BPMN models,  to provide a classification of BPMN collaborations.  In particular, we refer to collaborations involving processes with arbitrary topology, thus overcoming the well-structuredness limitations. 
The proposed classification is based on some of the most important correctness properties in the business process domain, namely safeness and soundness. We prove, with a uniform formal framework, some  conjectured and expected results and, most of all, we achieve novel results for BPMN collaborations concerning the relationships between safeness and soundness, and their compositionality, that represent major advances in the state-of-the-art.
\end{abstract}

\input{Introduction}

\input{backgroud}

\input{classification}

\input{formal}

\input{wellstructure}

\input{relationsonproerties}

\input{compositionality}
\input{Related}

\input{conclusions}
\bibliographystyle{eptcs}
\bibliography{bibclassificationpaper}
\end{document}

%% file: MacroClassification.tex
\newcommand{\bpmn}{BPMN}

\newcommand{\collabs}{C}
\newcommand{\procs}{P}

\newcommand{\poolG}[2]{{\sf pool}(#1,#2)}

\newcommand{\edgeSetE}{\mathbb{E}}
\newcommand{\edgename}{{\sf e}}
\newcommand{\edgeList}{E}

\newcommand{\msgSetM}{\mathbb{M}}

\newcommand{\messagename}{{\sf m}}

\newcommand{\isWS}[1]{\mathit{isWS}(#1)}

\newcommand{\evStartG}[2]{{\sf start}(#1, #2)}
\newcommand{\evEndG}[2]{{\sf end}(#1, #2)}
\newcommand{\evTerminateG}[1]{{\sf terminate}(#1)}
\newcommand{\emptyG}[2]{{\sf empty}(#1,#2)}

\newcommand{\edges}[1]{{\sf edges}(#1)}

\newcommand{\evStartRcvG}[3]{{\sf startRcv}(#1,#2, #3)}
\newcommand{\evEndSndG}[3]{{\sf endSnd}(#1,#2,#3)}
\newcommand{\evInterRcvG}[3]{{\sf interRcv}(#1,#3, #2)}
\newcommand{\evInterSndG}[3]{{\sf interSnd}(#1,#3, #2)}
\newcommand{\andSplitG}[2]{{\sf andSplit}(#1,#2)}
\newcommand{\andJoinG}[2]{{\sf andJoin}(#1,#2)}
\newcommand{\xorSplitG}[2]{{\sf xorSplit}(#1,#2)}
\newcommand{\xorJoinG}[2]{{\sf xorJoin}(#1,#2)}

\newcommand{\taskG}[2]{{\sf task}(#1,#2)}
\newcommand{\tasksendG}[3]{{\sf taskSnd}(#1,#3,#2)}
\newcommand{\taskreceiveG}[3]{{\sf taskRcv}(#1,#3,#2)}
\newcommand{\eventbasedG}[2]{{\sf eventBased}(#1,#2)}
\newcommand{\subprocG}[3]{{\sf subProc}(#1,#2,#3)}

\newcommand{\transitionColl}[2]{\xrightarrow[#2]{#1}}

\newcommand{\edgeG}[1]{#1}
\newcommand{\incToken}[2]{inc(#1,#2)}
\newcommand{\decToken}[2]{dec(#1,#2)}

\newcommand{\marked}[2]{marked(#1,#2)}

\newcommand{\stateedge}{\sigma}
\newcommand{\statemsg}{\delta}

\newcommand{\statefunc}[1]{\sigma(#1)}
\newcommand{\statefuncmsg}[1]{\delta(#1)}

\newcommand{\internalAction}{\tau}

\newcommand{\token}{{\sf n}}
\newcommand{\collabspar}{\mid\!\mid}
\newcommand{\procspar}{\mid\!\mid}

\newcommand{\orgname}{{\sf p}}
 \newcommand{\isInit}[1]{\mathit{isInit}(#1)}

\newcommand{\getCmpP}[1]{\mathit{end}(#1)}

\newcommand{\isZero}[1]{\mathit{isZero}(#1)}
 
  \newcommand{\maxMarking}[1]{\mathit{maxMarking}(#1)}

\newcommand{\poolLabel}{\alpha}
\newcommand{\procLabel}{\alpha}

\newcommand{\receiveAction}[1]{?#1}
\newcommand{\morte}{kill}
\newcommand{\intAction}{\epsilon}

\newenvironment{proofS}
{
        \paragraph{Proof (sketch).}
        }
 {}

%% file: annotationnote.tex
\usepackage{ifthen}
\usepackage{pifont}
\usepackage{amssymb}

\newboolean{showtodos}
\setboolean{showtodos}{true} % toggle to show or hide comments
\ifthenelse{\boolean{showtodos}}
  {
    \newcommand{\todomarker}[4]{\marginpar{\textcolor{#1}{#2}}\footnote{
    \textcolor{#1}{\it\scriptsize {\textbf{\sf \underline{#3}:~#4}}}}}
  }
  {
    \newcommand{\todomarker}[4]{}
  }

\newboolean{showcomments}
\setboolean{showcomments}{true} % toggle to show or hide comments
\ifthenelse{\boolean{showcomments}}
  {
    \newcommand{\notemarker}[4]{{\textcolor{#1}{#2}}\footnote{
    \textcolor{#1}{\it\scriptsize {\textbf{\sf \underline{#3}:~#4}}}}}
  }
  {
    \newcommand{\notemarker}[4]{}
  }

%% file: macroProoftree.tex
\newdimen\proofrulebreadth \proofrulebreadth=.04em
\newdimen\proofdotseparation \proofdotseparation=1.25ex
\newdimen\proofrulebaseline \proofrulebaseline=2ex
\newcount\proofdotnumber \proofdotnumber=3
\let\then\relax
\def\hfi{\hskip0pt plus.0001fil}
\mathchardef\squigto="3A3B
\newif\ifinsideprooftree\insideprooftreefalse
\newif\ifonleftofproofrule\onleftofproofrulefalse
\newif\ifproofdots\proofdotsfalse
\newif\ifdoubleproof\doubleprooffalse
\let\wereinproofbit\relax
\newdimen\shortenproofleft
\newdimen\shortenproofright
\newdimen\proofbelowshift
\newbox\proofabove
\newbox\proofbelow
\newbox\proofrulename
\def\shiftproofbelow{\let\next\relax\afterassignment\setshiftproofbelow\dimen0 }
\def\shiftproofbelowneg{\def\next{\multiply\dimen0 by-1 }%
\afterassignment\setshiftproofbelow\dimen0 }
\def\setshiftproofbelow{\next\proofbelowshift=\dimen0 }
\def\setproofrulebreadth{\proofrulebreadth}

\def\prooftree{% NESTED ZERO (\ifonleftofproofrule)
\ifnum  \lastpenalty=1 \then   \unpenalty \else
\onleftofproofrulefalse \fi
\ifonleftofproofrule \else   \ifinsideprooftree
        \then   \hskip.5em plus1fil
        \fi
\fi
\bgroup% NESTED ONE (\proofbelow, \proofrulename, \proofabove,
\setbox\proofbelow=\hbox{}\setbox\proofrulename=\hbox{}%
\let\justifies\proofover\let\leadsto\proofoverdots\let\Justifies\proofoverdbl
\let\using\proofusing\let\[\prooftree
\ifinsideprooftree\let\]\endprooftree\fi
\proofdotsfalse\doubleprooffalse
\let\thickness\setproofrulebreadth
\let\shiftright\shiftproofbelow \let\shift\shiftproofbelow
\let\shiftleft\shiftproofbelowneg
\let\ifwasinsideprooftree\ifinsideprooftree
\insideprooftreetrue
\setbox\proofabove=\hbox\bgroup$\displaystyle % NESTED TWO
\let\wereinproofbit\prooftree
\shortenproofleft=0pt \shortenproofright=0pt \proofbelowshift=0pt
\onleftofproofruletrue\penalty1 }

\def\eproofbit{% NESTED TWO
\ifx    \wereinproofbit\prooftree \then   \ifcase \lastpenalty
        \then   \shortenproofright=0pt  % 0: some other object, no indentation
        \or     \unpenalty\hfil         % 1: empty hypotheses, just glue
        \or     \unpenalty\unskip       % 2: just had a tree, remove glue
        \else   \shortenproofright=0pt  % eh?
        \fi
\fi
\global\dimen0=\shortenproofleft \global\dimen1=\shortenproofright
\global\dimen2=\proofrulebreadth \global\dimen3=\proofbelowshift
\global\dimen4=\proofdotseparation
$\egroup  % NESTED ONE
\shortenproofleft=\dimen0 \shortenproofright=\dimen1
\proofrulebreadth=\dimen2 \proofbelowshift=\dimen3
\proofdotseparation=\dimen4
}

\def\proofover{% NESTED TWO
\eproofbit % NESTED ONE
\setbox\proofbelow=\hbox\bgroup % NESTED TWO
\let\wereinproofbit\proofover
$\displaystyle
}%
\def\proofoverdbl{% NESTED TWO
\eproofbit % NESTED ONE
\doubleprooftrue
\setbox\proofbelow=\hbox\bgroup % NESTED TWO
\let\wereinproofbit\proofoverdbl
$\displaystyle
}%
\def\proofoverdots{% NESTED TWO
\eproofbit % NESTED ONE
\proofdotstrue
\setbox\proofbelow=\hbox\bgroup % NESTED TWO
\let\wereinproofbit\proofoverdots
$\displaystyle
}%
\def\proofusing{% NESTED TWO
\eproofbit % NESTED ONE
\setbox\proofrulename=\hbox\bgroup % NESTED TWO
\let\wereinproofbit\proofusing
\kern0.3em$ }

\def\endprooftree{% NESTED TWO
\eproofbit % NESTED ONE
  \dimen5 =0pt% spread of hypotheses
\dimen0=\wd\proofabove \advance\dimen0-\shortenproofleft
\advance\dimen0-\shortenproofright
\dimen1=.5\dimen0 \advance\dimen1-.5\wd\proofbelow \dimen4=\dimen1
\advance\dimen1\proofbelowshift \advance\dimen4-\proofbelowshift
\ifdim  \dimen1<0pt \then   \advance\shortenproofleft\dimen1
        \advance\dimen0-\dimen1
        \dimen1=0pt
        \ifdim  \shortenproofleft<0pt
        \then   \setbox\proofabove=\hbox{%
                        \kern-\shortenproofleft\unhbox\proofabove}%
                \shortenproofleft=0pt
        \fi
\fi
\ifdim  \dimen4<0pt \then   \advance\shortenproofright\dimen4
        \advance\dimen0-\dimen4
        \dimen4=0pt
\fi
\ifdim  \shortenproofright<\wd\proofrulename \then
\shortenproofright=\wd\proofrulename \fi
\dimen2=\shortenproofleft \advance\dimen2 by\dimen1
\dimen3=\shortenproofright\advance\dimen3 by\dimen4
\ifproofdots \then
        \dimen6=\shortenproofleft \advance\dimen6 .5\dimen0
        \setbox1=\vbox to\proofdotseparation{\vss\hbox{$\cdot$}\vss}
        \setbox0=\hbox{%
                \kern\dimen6
                $\vcenter to\proofdotnumber\proofdotseparation
                        {\leaders\box1\vfill}$%
                \unhbox\proofrulename}%
\else   \dimen6=\fontdimen22\the\textfont2 % height of maths axis
        \dimen7=\dimen6
        \advance\dimen6by.5\proofrulebreadth
        \advance\dimen7by-.5\proofrulebreadth
        \setbox0=\hbox{%
                \kern\shortenproofleft
                \ifdoubleproof
                \then   \hbox to\dimen0{%
                        $\mathsurround0pt\mathord=\mkern-6mu%
                        \cleaders\hbox{$\mkern-2mu=\mkern-2mu$}\hfill
                        \mkern-6mu\mathord=$}%
                \else   \vrule height\dimen6 depth-\dimen7 width\dimen0
                \fi
                \unhbox\proofrulename}%
        \ht0=\dimen6 \dp0=-\dimen7
\fi
\let\doll\relax
\ifwasinsideprooftree \then   \let\VBOX\vbox \else
\ifmmode\else$\let\doll=$\fi
        \let\VBOX\vcenter
\fi
\VBOX   {\baselineskip\proofrulebaseline \lineskip.2ex
        \expandafter\lineskiplimit\ifproofdots0ex\else-0.6ex\fi
        \hbox   spread\dimen5   {\hfi\unhbox\proofabove\hfi}%
        \hbox{\box0}%
        \hbox   {\kern\dimen2 \box\proofbelow}}\doll%
\global\dimen2=\dimen2 \global\dimen3=\dimen3
\egroup % NESTED ZERO
\ifonleftofproofrule \then   \shortenproofleft=\dimen2 \fi
\shortenproofright=\dimen3
\onleftofproofrulefalse \ifinsideprooftree \then   \hskip.5em plus
1fil \penalty2 \fi }

%% file: Introduction.tex
%!TEX root = ../Main_classification_Express.tex
\section{Introduction} 
Modern organisations recognise the importance of having tools to describe how to behave in
order to reach their own objectives. This is generally reflected in a business process model that is characterised as ``\textit{a collection of related and structured activities undertaken by one or more organisations in order to pursue some particular   goal.   [\ldots]   Business processes are often interrelated since the execution of a business process often results in the activation of related  business processes within the same or other organisations}'' \cite{lindsaybusiness2003}.
Up to now, several languages have been proposed to represent business process models. The OMG standard \bpmn{} \cite{omgbusiness2011} is the most prominent language, since it is widely accepted in both academia and industry. In particular, \bpmn{} collaboration models are used to describe distributed and complex scenarios where multiple participants interact via the exchange of messages.  

Eventhough widely accepted, \bpmn{}'s major drawback is related to the possible misunderstanding of its execution semantics, defined by means of natural text descriptions, sometimes containing misleading information \cite{sucheniaselected2017}. 
To overcome this issue, much effort has been devoted to formalise \bpmn{} semantics by means of mapping it to other formal languages. The most relevant is the one to Petri Nets provided by Dijkman et al. \cite{dijkmansemantics2008}.
However, models resulting from a mapping inherit constraints given by the target formal language and so far none of them considers \bpmn{} features such as the different abstraction levels (i.e., collaboration, process, and sub-process), the asynchronous communication model, and the notion of termination due to different types of end event (i.e., simple, message throwing, and terminate). 
 
Our investigation is based on a formal characterisation of the \bpmn{} semantics specifically given for collaboration models. It is used to formally define a classification of these collaboration models according to relevant properties of the business process domain.  It is worth noticing that our work aims at providing a classification specific for the \bpmn{} notation.  To this aim, our formal semantics is directly defined on \bpmn{} elements.  
Our intention is to introduce a unique formal framework to allow \bpmn{} designers to achieve a better understanding of their models, and relative properties. This results in a systematic methodological approach to improve the design of \bpmn{} collaborations. % during their lifecycle. 

As a distinctive aspect, the proposed semantics supports models with arbitrary topology, including of course also `well-structured' ones~\cite{kiepuszewskistructured2000}.
This is necessary to enable a classification of both structured and unstructured models with respect to specific properties. Unstructured models can often be studied by means of structured ones at the cost of considerable increasing their size \cite{dumasunderstanding2012}, altough this is not always possible~\cite{polyvyanyystructuring2012,polyvyanyymaximal2014}. 
Looking at the public repository of \bpmn{} models provided by the BPM Academic Initiative (\url{http://bpmai.org}) we noticed that unstructured models are largely used in practice, especially when the size of the models is significant \cite{TR}.

Regarding the considered properties, our classification relies on a well-known class of properties in the domain of business process management, namely \emph{safeness} \cite{vanderaalstworkflow2000} and \emph{soundness} \cite{vanderaalstsoundness2011,vanderaalstprocessoriented1999}. 
So far, no formal definition of such properties directly given on \bpmn{} is provided, despite the large body of work on this topic. 
%This is our first contribution. 
%
We reconcile in our single framework properties taken into account by different languages, like 
Petri Nets \cite{muratapetri1989}, Workflow Nets \cite{vanderaalstworkflow2000},  and Elementary Nets \cite{rozenbergelementary1998}. Studying different properties in the same framework does not leave any room for ambiguity, and increases the potential for formal reasoning on \bpmn{}.
Differently from other formal notations, our framework primitively allows to express important features of the \bpmn{} �notation, such as message passing and relative soundness properties (e.g. {message-relaxed soundness}).  

Hence, the main contribution of the paper is a classification of \bpmn{} collaborations according to relevant properties of the domain. 
More in detail, we prove that a well-structured collaboration is always safe, but the reverse does not hold. Moreover, well-structuredness implies soundness only at the process level, while there are well-structured collaborations that are not sound. 
Regarding the relationships between soundness and safeness, we prove that soundness does not imply safeness. Indeed, there are unsafe models that are sound. Similarly, sound models are not necessarily safe. 
Moreover, we study safeness and soundness compositionality in the domain of business process management, and  show how specific \bpmn{} constructs, namely terminate  event and sub-processes, move certain BPMN models from one class to another. 
To illustrate both our formal framework and properties, as well as their relationships, we rely on a running example concerning a travel agency scenario.

The rest of the paper is organised as follows. Sec.~\ref{sec:Backgroud} provides background notions on \bpmn{} and the considered properties. Sec.~\ref{sec:BPMNclassification}  introduces a first insight into the obtained results, while Sec.~\ref{sec:formal} introduces the proposed formal framework. Sec.~\ref{sec:Definitions} provides the definition of properties, and Sec.~\ref{sec:rela} makes it clear the relationships between them. Sec.~\ref{sec:compostionality} presents the study on safeness and soundness compositionality. Finally, Sec.~\ref{sec:Related} discusses related works, and Sec.~\ref{sec:conclusion} concludes the paper.

%% file: backgroud.tex
%!TEX root = ../Main_classification_Express.tex
 
\section{Background Notions}
\label{sec:Backgroud}
In this section we first provide some basic notions on elements that can be included in \bpmn\ collaboration  diagrams. Then, we present a travel agency collaboration scenario, used throughout the paper as a  running example. Finally, we illustrate the properties we considered in the classification. 

\smallskip
\noindent
\textbf{Basic Notions on \bpmn{}.}
Here we do not aim at providing a complete presentation of the standard, but a discussion of the main concepts of BPMN~\cite{omgbusiness2011} we use in the following. 
Our choice of the considered \bpmn\ fragment is driven by practical aspects. Indeed, as shown in~\cite{muehlenhow2008}, even if the \bpmn\ specification is quite wide, only less than 20\% of its vocabulary is used regularly in designing business process models. 

\begin{figure}
\centering
 \includegraphics[width=\textwidth]{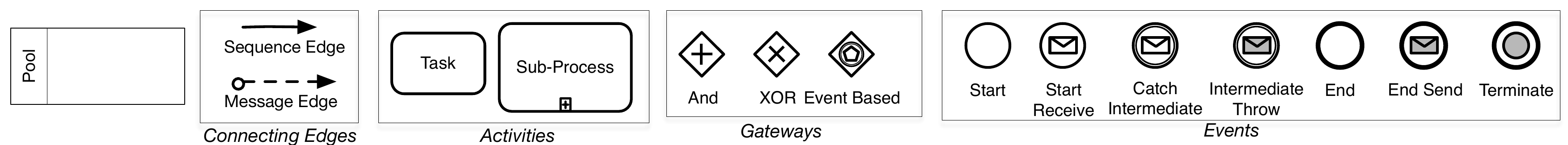}
   \vspace*{-.8cm}
  \caption{Considered BPMN 2.0 Elements.}
  \label{fig:bpmn}
  \vspace*{-.5cm}
\end{figure}

The \bpmn{} elements we considere are those constructs that are most used in practice. They are briefly described below and reported in Fig.~\ref{fig:bpmn}. 
 \textbf{Pools} are used to represent participants or organisations involved in a collaboration, and include details on internal process specifications and related elements.  
   \textbf{Connecting Edges} are used to connect process elements
  in the same or different pools.  \textit{Sequence Edge} is a solid connector used to specify the internal flow of the process, thus ordering elements in the same pool, while \textit{Message Edge} is a dashed connector
  used to visualise communication flows between organisations.
  \textbf{Activities} are used to represent specific works to be performed
  within a process.  In particular, a \textit{task}  is an atomic activity. It represents work that can not interrupted. It  can be also used to send and receive messages.  A \textit{sub-process} represents work that is broken down to a finer level of detail.
   \textbf{Gateways} are used to manage the control flow of a process both
  for parallel activities and choices. Gateways act as either join nodes (merging incoming sequence edges) or
  split nodes (forking into outgoing sequence edges).  Different types
  of gateways are available.
A \textit{XOR} gateway gives the possibility to model
    choices. In particular, a XOR-Split gateway is used after a
    decision to fork the flow into branches. When executed, it
    activates exactly one outgoing edge. A XOR-Join gateway acts as a
    pass-through, meaning that it is activated each time the gateway
    is reached. 
   An \textit{AND} gateway enables parallel execution flows. An
    AND-Split gateway is used to model the parallel execution of two
    or more branches, as all outgoing sequence edges are activated
    simultaneously. An AND-Join gateway synchronises the execution of
    two or more parallel branches, as it waits for all incoming
    sequence edges to complete before triggering the outgoing flow.
   An \textit{Event-Based} is used after a
    decision to fork the flow into branches according to external choice. Its outgoing branches activation depends on taking
    place of catching events. Basically, such events are in a race
    condition, where the first event that is triggered wins and
    disables the other ones. 
  \textbf{Events} are used to represent something that can
  happen. An event can be a \textit{Start Event} representing the
  point from which a process starts, an \textit{Intermediate Event}
  representing something that happens during process execution, or an
  \textit{End Event} representing the process termination. Events are
  drawn as circles.  When an event is source or target of a message
  edge, it is called \textit{Message Event}.
  %\footnote{In the standard
%    \cite{omgbusiness2011}, message events are represented by
%    circles enclosing an envelope, which is white in case of catching
%    events and black in case of throwing ones. However, the
%    information provided by the presence and the color of the envelope
%    is redundant, as the type of the event is made clear by the
%    possible connection with a message edge. Therefore, to keep the
%    notation clean, which is useful for our formalisation purpose, we
%    omit the envelope from message events without loss of clear
%    meaning.}.  
    According to the different kinds of message edge
  connections, we distinguish between: (i) \emph{Start Message Event} is a start event with an incoming
    message edge; the event element catches a message and starts a process; (ii) \emph{Throw Intermediate Event} is an intermediate event with
    an outgoing message edge; the event element sends a message; (iii)  \emph{Catch Intermediate Event} is an intermediate event with
    an incoming message edge; the event element receives a message, (iv) \emph{End Message Event} is an end event with an outgoing
    message edge; the event element sends a message and ends the 
    process.
  We also refer to a particular type of end event, the
  \textit{Terminate End Event} able to stop and abort the running
  process.
Finally, a key concept related to the \bpmn{} process execution refers to the notion of 
\emph{token}. The \bpmn{} standard states that ``a token is a
theoretical concept that is used as an aid to define the behaviour of a
process that is being
performed'' \cite[Sec. 7.1.1]{omgbusiness2011}. A token is
commonly generated by a start event, traverses the sequence edges of
the process and passes through its elements 
enabling their execution, and it is consumed by an end event when process completes. 
The distribution of tokens in the process elements is called
\emph{marking}, therefore the \emph{process execution} is defined in terms
of marking evolution. In the collaboration, the process execution also triggers message flow able to genere messages. We will refer them as message flow token.

 \smallskip
\noindent
\textbf{Travel Agency Collaboration Scenario.}
\begin{figure}[h]
  \centering
    %\vspace{-.9cm}
 \includegraphics[width=\textwidth]{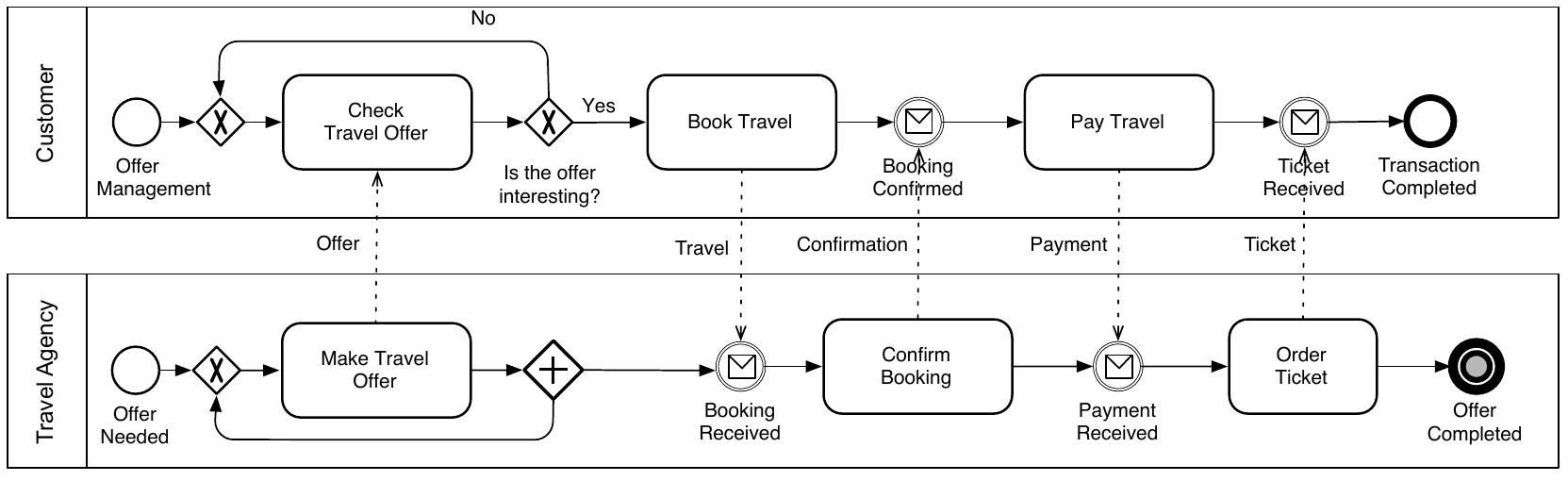}
    \vspace{-.4cm}
\caption{\bpmn\ collaboration diagram of a Travel Agency scenario.}
\vspace{-.5cm}
\label{fig:runexample}
\end{figure}
Considering some of the introduced \bpmn{} elements we can obtain the travel agency scenario in Fig.~\ref{fig:runexample}, which combines in a collaboration model the activities of a Customer and a Travel Agency. 
In particular, the Travel Agency continuously offers travels to the Customer, until an offer is accepted. If the Customer is interested in one offer, she decides to book the travel and refuses all the other offers that the Travel Agency insistently proposes. As soon as the booking is received by the Travel Agency, it sends back a confirmation message, and asks for the payment of the travel. When this is completed the ticket is sent to the Customer, and the Travel Agency activities end.
The processes of the Customer and of the Travel Agency are represented inside two \emph{pools}.  
Considering the Customer pool, from left to right, we have that as soon as the process starts, due to its \emph{start event}, the Customer checks a received travel offer. This is done by executing a \emph{receiving task}. Then, she decides either to book the travel or to wait for other offers, by means of a loop composed of two \emph{gateways}: a XOR-Join and a XOR-Split. After the Customer finds the interesting offer, she books the travel, by sending a message to the Travel Agency via  a \emph{sending task}, and waits for the booking confirmation. As soon as she receives the booking confirmation, through an \emph{intermediate receiving event}, she pays the travel, receives the ticket from the Travel Agency and her specific works terminate by means of an \emph{end event}.
Considering the work of the Travel Agency, as soon as its process starts, it makes travel offers to the Customer, by means of a \emph{sending task}. Thanks to the looping behaviour obtained by an AND-Split combined with a XOR-Join, it continuously makes offers. At the same time, it proceeds in order to receive a booking via   an \emph{intermediate receiving event}. Then, it confirms the booking and sends a notification to the Customer. Finally, after receiving the payment, it orders and sends the ticket, thus completing its activities by means of a \emph{terminate event},  which stops and aborts the running process, including the offering of travels.

\smallskip
\noindent
\textbf{Well-structuredness, Safeness and Soundness for \bpmn{}.}
We introduce now an informal definition of \emph{well-structuredness}, \emph{safeness} and \emph{soundness} for \bpmn{} models, while their formalisation is provided in Section~\ref{sec:Definitions}. 
In particular, well-structuredness relates to the way the various model elements are connected with each other, while safeness and soundness relate to the way a process model can be executed. 

A \bpmn{} process model is \emph{well-structured}~\cite{kiepuszewskistructured2000}, if for every split gateway there is a corresponding join gateway such that the fragment of the model between the split and the join forms a single-entry-single-exit process component. 
%As an example, the process in Fig.~\ref{fig:WS} is the well-structured version of the unstructured process in Fig.~\ref{fig:nonWS}.
%
%\begin{figure}[t]
%\centering
%\begin{minipage}[c]{.50\textwidth}
%%\vspace{-0.8cm}
%\centering\setlength{\captionmargin}{0pt}%
%  \includegraphics[width=0.8\textwidth]{img/WS.pdf}
%    \vspace{-.1cm}
%  \caption{A WS Process Model. \label{fig:WS}}
%\end{minipage}%
%\hspace{-2mm}%
%\begin{minipage}[c]{.50\textwidth}
%%\vspace{-1cm}
%\centering\setlength{\captionmargin}{0pt}%
%  \includegraphics[width=0.8\textwidth]{img/WSeNoNWS.pdf}
%    \vspace{-.3cm}
%  \caption{A non WS Process Model.     \label{fig:nonWS}}
%\end{minipage}
%   \vspace{-.1cm}
%\end{figure}
The notion of well-structuredness can be extended to process collaborations (see Def. \ref{WSC}), requiring that  the processes of all involved organisations are well-structured.

A \bpmn{} process model is \emph{safe}\footnote{Notably, the notion of safeness is different from that of safety, since it is a standard term in the BPM literature.}, if during its execution no more than one token occurs along the same sequence edge (see Def. \ref{def:safenessP}).  This property is inspired by the Petri-Net literature, where safeness means that a Petri Net does not contain more than one token in all reachable markings \cite{vanderaalstworkflow2000}.
Safeness naturally extends to process collaborations, requiring that the processes of all involved organisations satisfy it considering the overall collaboration execution (see Def. \ref{def:safenessC}).

A \bpmn{} process model is \emph{sound}, if it can complete its execution without leaving active elements and all the model elements can be activated in at least one of the execution traces (see Def. \ref{def:soundnessProc}).
%Generally, ``a business process model is sound if it
%can successfully terminate without lefting over active elements and all the model elements can
%be activated in one of the execution traces'' \cite{el-saber_cmmi-cm_2015}.
This property is inspired by the business process literature that, since the mid nineties, presents several versions of soundness considering different modelling languages  
\cite{vanderaalstworkflow2000,vanderaalstsoundness2011,vanderaalstprocessoriented1999,el-sabercmmi-cm2015}. 
To escape from this jungle of definitions and focussing on the specificities of \bpmn{}, here 
we based on the classical soundness definition for processes \cite{DBLP:books/sp/DumasRMR18} that informally requires the satisfaction of three sub-properties:
\emph{(i)} \emph{Option to complete}: any running process instance must eventually complete,
\emph{(ii)} \emph{Proper completion}:  at the moment of completion, each token of the process
instance must be in a different end event,
\emph{(iii)}  \emph{No dead activities}:  any activity can be executed in at least one process instance.
Considering the \bpmn\ notion of completion \cite[pp. 426, 431]{omgbusiness2011}, requiring that all tokens in that instance must reach an end node, the \emph{no dead activities} property is equivalent to requiring the complete execution of each process instance. In fact, the only way to have dead activities is that the incoming sequence flow of that activity is never reached by a token. This can happen either when there is a deadlock upstream the considered activity or when there are some conditions on gateways. The first case is subsumed in the notion of completion, while the second case is not caught by our semantics since reasoning on models with data is out of the scope of this work. 
Finally, we extend the soundness property to process collaborations (see Def.  \ref{def:soundnessC}).

%% file: classification.tex
%!TEX root = ../Main_classification_Express.tex
\section{Classification Results}
\label{sec:BPMNclassification}

\begin{figure*}[h]
  \centering
    \vspace{-0.2cm}
  \includegraphics[width=0.6\textwidth]{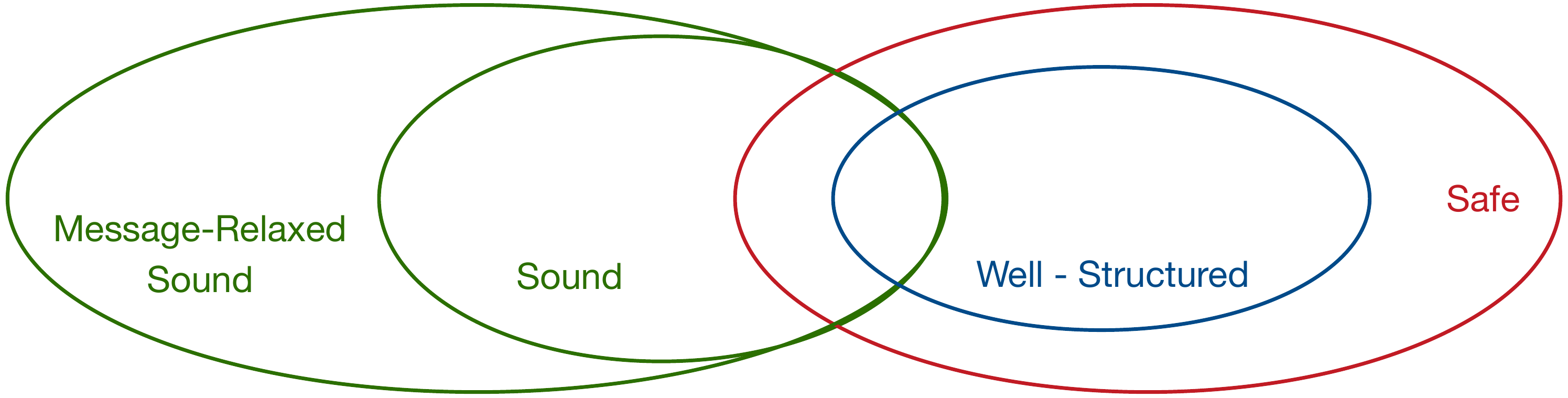}
    \vspace{-.2cm}
  \caption{Classification of BPMN collaborations.}
  \vspace{-.4cm}
  \label{fig:setsclass}
\end{figure*}

In this section, we show how \bpmn{} collaborations can be classified according to well-structuredness, safeness, and soundness. Differently from other classifications 
\cite{dehnertsuitability2005,vanderaalststructural1996} reasoning only at the process level and considering the Workflow Nets semantics, our study 
directly addresses \bpmn{} collaboration models. The obtained results we formally prove are synthesised in the Euler diagram in Fig.~\ref{fig:setsclass}, showing in particular that:
 \begin{itemize} 
\item[\emph{(i)}] all well-structured collaborations are safe, but the reverse does not hold;
\item[\emph{(ii)}] there are well-structured collaborations that are neither sound  nor message-relaxed sound;
\item[\emph{(iii)}] there are sound and message-relaxed sound collaborations that are not safe.
\end{itemize}

\smallskip
\noindent
\textbf{Advances with respect to already available classifications.}
Result \emph{(i)} demonstrates that well-structured collaborations represent a subclass of safe collaborations. We also show that such a relation is valid at the process level, where the classification relaxes the existing results on Workflow Nets, stating that a process model to be safe has to be not only well-structured, but also sound \cite{vanderaalststructural1996}. 
Result \emph{(ii)} shows that there are well-structured collaborations that are not sound. This is also valid at  process level, confirming results provided on Workflow Nets, where well-structuredness implies soundness \cite{van2006history}, and relaxing the one obtained in Petri Nets  \cite{dehnertsuitability2005}, where relaxed soundness and well-structuredness together imply soundness.
Results \emph{(i)} and \emph{(ii)} together confirm limits of well-structuredness as a correctness criterion. Indeed,  considering only well-structuredness is too strict, as some safe and sound models that are not well-structured result discarded right from the start.
Result \emph{(iii)} shows that there are sound and message-relaxed sound collaborations that are not safe. This also can be observed at the process level resulting in a novel contribution strictly related to the expressiveness of \bpmn\ and its differences with respect to other workflow languages.
In fact, Van der Aalst shows that soundness of a Workflow Net
is equivalent to liveness and boundedness of the corresponding short-circuited Petri Net \cite{vanderaalstverification1997}. 
Similarly, in workflow graphs and, equivalently, free-choice Petri Nets, soundness can be characterized in terms of two types of local errors, viz. deadlock and lack of synchronization: a workflow graph is sound if it contains neither a deadlock nor a lack of synchronization \cite{favresymbolic2010,prinzfast2013}.
Thus, a sound workflow is always safe. In \bpmn\ instead there are unsafe processes that are sound.
Summing up,  our results %\emph{(i)} together with result \emph{(ii)} and \emph{(iii)}  
are novel and also influence the reasoning at process level. This is mainly due to the effects of the terminate event and sub-process behaviour, that impact on the classification of models both at the process and  collaborations levels, as shown in the following.

\smallskip
\noindent
\textbf{Advances in Classifying \bpmn{} Models.} 
Our \bpmn{}  formalisation considers as first-class citizens \bpmn{} specificities,
reasoning on: collaboration, process and sub-process levels;  asynchronous communication; 
and the \bpmn{} completion notion that distinguishes the effect of end and terminate events. 

Considering collaboration models, we can observe pools that exchange message tokens, while in each pool the execution is rendered by the movements of the sequence flow  
tokens at process level. In this setting, there is a clear difference between the notion of safeness directly defined on BPMN collaborations compared to the one defined on Petri Nets and applied to the Petri Nets resulting from the translation of BPMN collaborations,
%According to known mappings mapping, 
e.g. via the mappings in \cite{dijkmansemantics2008} and in  \cite{KunzeW16}. Safeness of a BPMN collaboration only refers to tokens on the sequence edges of the involved processes, while in its Petri Nets translation refers to tokens both on message and sequence edges. Such distinction is not considered in the  mappings, because a message is rendered as a (standard) token in a place.  Hence, a safe BPMN collaboration where the same message is sent more than once (e.g., via a loop), it is erroneously considered unsafe by relying on the Petri Nets notion (i.e., 1-boundedness), because enqueued messages are rendered as a place with more than one token. Therefore, the notion of safeness defined for Petri Nets cannot be safely applied as it is to collaboration models.
Similarly, regarding the soundness property, we are able to consider different notions of soundness according to the requirements we impose on message queues (e.g., ignoring or not the pending messages). Again, due to lack of distinction between message and sequence edges, these fine-grained reasoning are precluded using the current translations from \bpmn{} to Petri Nets.

The study of \bpmn{} models via the frameworks based on Petri Nets has another limitation, concerning the management of the terminate event. 
Most of the available mappings, such as the ones in \cite{KunzeW16} and \cite{kheldounformal2017}, do not consider the termination end event, while in the one provided in \cite{dijkmansemantics2008} terminate events are treated as a special type of error events, which however occur mainly on sub-processes, whose translation assumes safeness.
This does not allow reasoning on most models that include the terminate event, and more in general on all models including unsafe sub-processes. 
Nevertheless, given the local nature of Petri Nets transitions, such cancellation patterns are difficult to handle. This is confirmed in \cite{ter2002workflow}, stating that modelling a vacuum cleaner, i.e., a construct to remove all the tokens from a given fragment of a net, is possible but results in a spaghetti-like diagram.

\begin{figure}[t]
\centering
\begin{tabular}{c@{\hspace*{1cm}}c}
  \includegraphics[width=.3\textwidth]{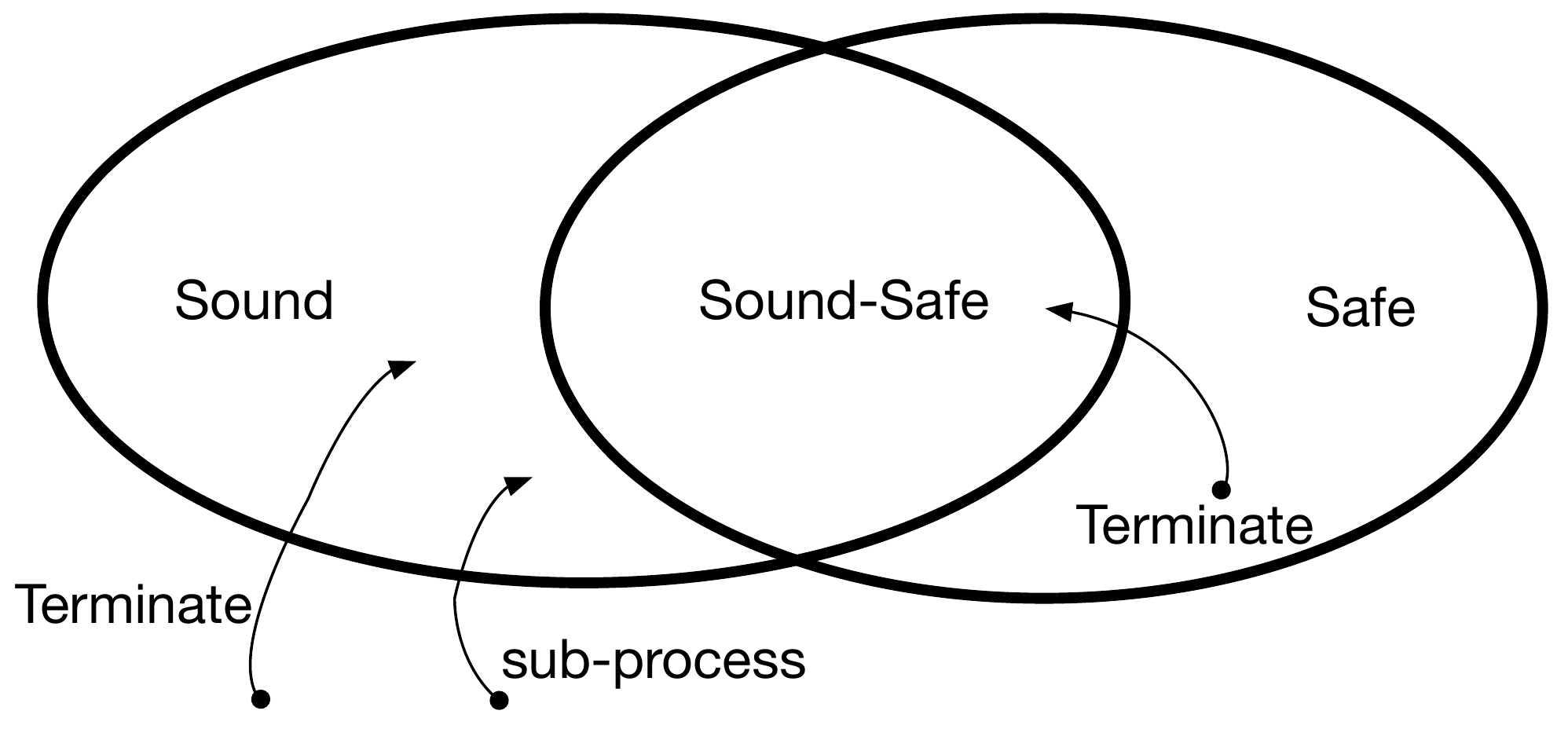}
&
  \includegraphics[width=.3\textwidth]{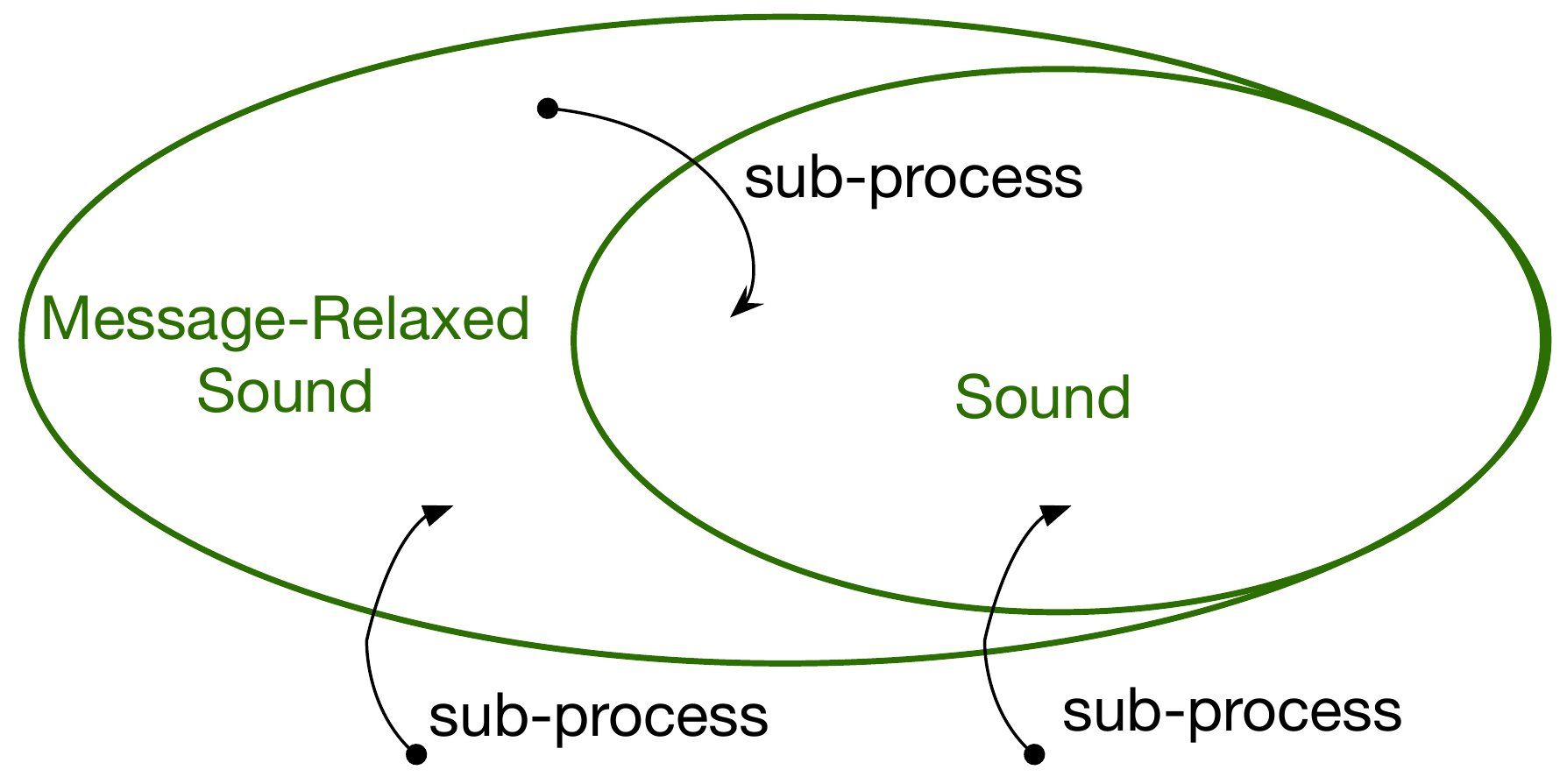}
\\[-.2cm]
(a) & (b)
\\
\end{tabular}
    \vspace{-.3cm}
  \caption{Reasoning at process (a) and collaboration (b) level. 
  \label{fig:kill}
  \label{fig:collsubproceffect}}
    \vspace{-.2cm}
\end{figure}

The ability of our formal framework to properly classify \bpmn{} models both at process and collaboration level, together with our treatment of the terminate event and sub-processes without any of the restrictions mentioned above, has led us to provide a more precise classification as synthesised by the Euler diagrams in Fig.~\ref{fig:kill}(a)  and Fig.~\ref{fig:collsubproceffect}(b).

\begin{figure}[t]
\centering
\begin{minipage}[c]{.38\textwidth}
%\vspace{-0.8cm}
\centering\setlength{\captionmargin}{0pt}%
  \includegraphics[width=0.9\textwidth]{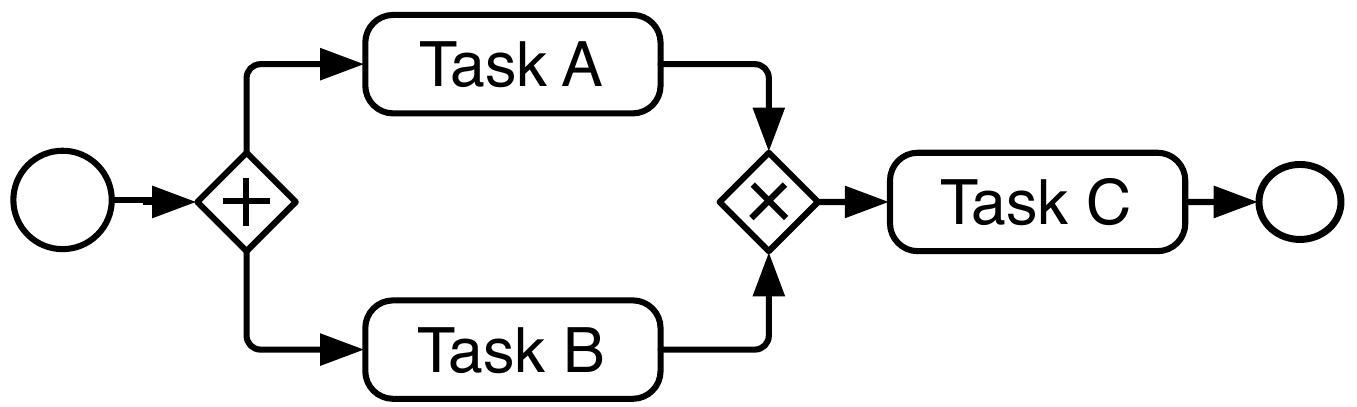}
  \vspace{-.3cm}
  \caption{Unsound process. \label{fig:soundmotivation}}
\end{minipage}%
\hspace{1mm}%
\begin{minipage}[c]{.6\textwidth}
%\vspace{-1cm}
\centering\setlength{\captionmargin}{0pt}%
  \includegraphics[width=.7\textwidth]{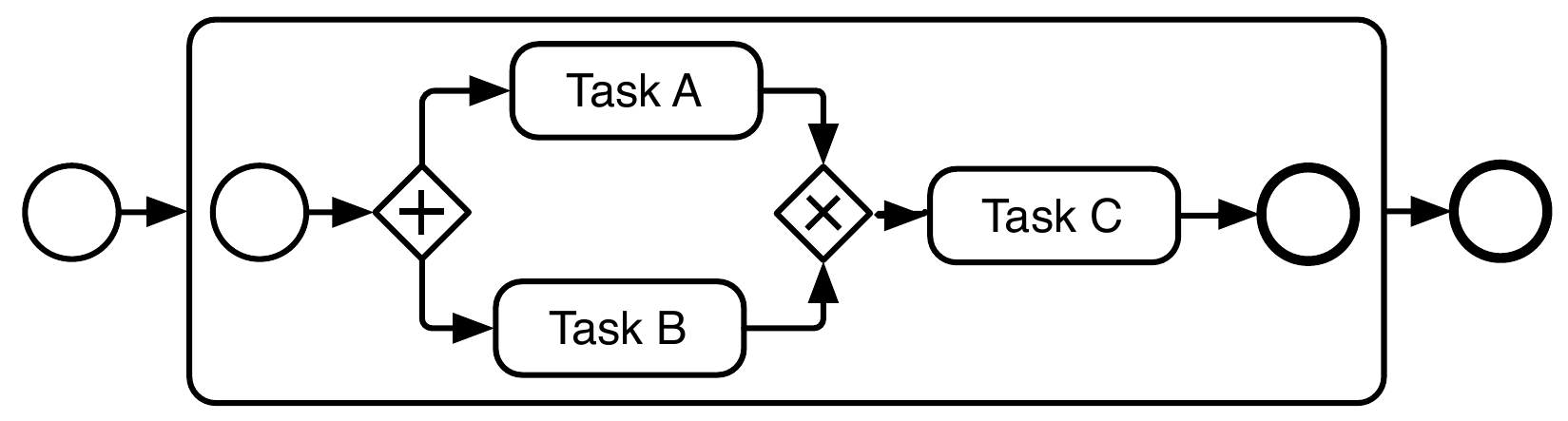}
      \vspace{-.3cm}
  \caption{Sound process with an unsound sub-process.  \label{fig:subproceffect}}
\end{minipage}
    \vspace{-0.7cm}
\end{figure}

In particular, Fig.~\ref{fig:kill}(a) depicts the reasoning that can be done at process level on soundness. 
%This is valid independently from safeness and well-structureness.
%
%
Here it emerges how the terminate event can affect model soundness, as using it in place of an end event may render a model that was unsound sound. 
For example, let us consider the process in Fig. \ref{fig:soundmotivation}; it is a simple process that 
first runs in parallel Task A and Task B, resulting in two possible executions of Task C, and finally completes with an end event. According to the proposed classification the model is unsound.
In fact,  from any reachable configuration of the process it is not possible to arrive in a (completed) configuration where all marked end events are marked exactly by a single token and all sequence edges are unmarked. 
%\begin{figure*}[htbp]
%  \centering
%    \vspace{-0.2cm}
%  \includegraphics[width=0.4\textwidth]{img/TokenGeneratorWithendEvent.pdf}
%    \vspace{-.2cm}
%  \caption{An example of sound process.}
%  \vspace{-.2cm}
%  \label{fig:soundmotivation}
%\end{figure*}
%
%
%
%
%
%
%,  due to the presence of more than one token possibly reaching the end event at different time, that is when the first token arrives there is still a token on the way.
%
%
%Now, let us modify the model by replacing the end event with a terminate end event. In such a case the process will even experience a sound scenario since
Now, let us consider another model, obtained from the one in Fig. \ref{fig:soundmotivation} by replacing the end event with a terminate event. The resulting model is sound. 
This is due to the behaviour of the terminate event that, when reached, removes all  tokens in the process. 
We underline that, although the two models are quite similar, in terms of our classification they result to be significantly different. 
Also the use of sub-processes can impact on the satisfaction of the soundness property. Fig.~\ref{fig:subproceffect} shows a simple process model where the unsound process in Fig. \ref{fig:soundmotivation}  is included in the sub-process.
%
%\begin{figure*}[htbp]
%  \centering
%    %\vspace{-1cm}
%  \includegraphics[width=0.7\textwidth]{img/SubOnclassification.pdf}
%    \vspace{-.2cm}
%  \caption{An example of sound process including sub-process.}
%  %\vspace{-.8cm}
%  \label{fig:subproceffect}
%\end{figure*}   
%
Notably, a sub-process is not syntactic sugar. % that can be removed via a sort of macro expansion. 
%
%Indeed, the  behaviour of the flattened process is different from that of the original process. derived by flattening it, i.e. by considering replacing the  At a first glance it could seem possible to consider the model unsound flatten the internal details of a sub-process within the rest of the model. This is not possible since in BPMN a sub-process is equipped with its own semantics. In particular, 
%
According to the \bpmn\ standard, a sub-process completes only when all the internal tokens are consumed, and then just one token is propagated along the including process  \cite{omgbusiness2011}. 
Thus,  it results that the model in Fig.~\ref{fig:subproceffect} is sound. Its behaviour would not correspond to that of the process obtained by flattening it, as the resulting model could be unsound. 
Notice, this reasoning is not affected by safeness and in particular, it cannot be extended to collaborations. In fact, as we discuss in Sec.~\ref{sec:compostionality}, when we compose two sound processes the resulting collaboration could be either sound or unsound.

\begin{figure}[h]
\centering
\vspace{-0.1cm}
\begin{minipage}[c]{.33\textwidth}
%\vspace{-0.8cm}
\centering\setlength{\captionmargin}{0pt}%
  \includegraphics[width=0.7\textwidth]{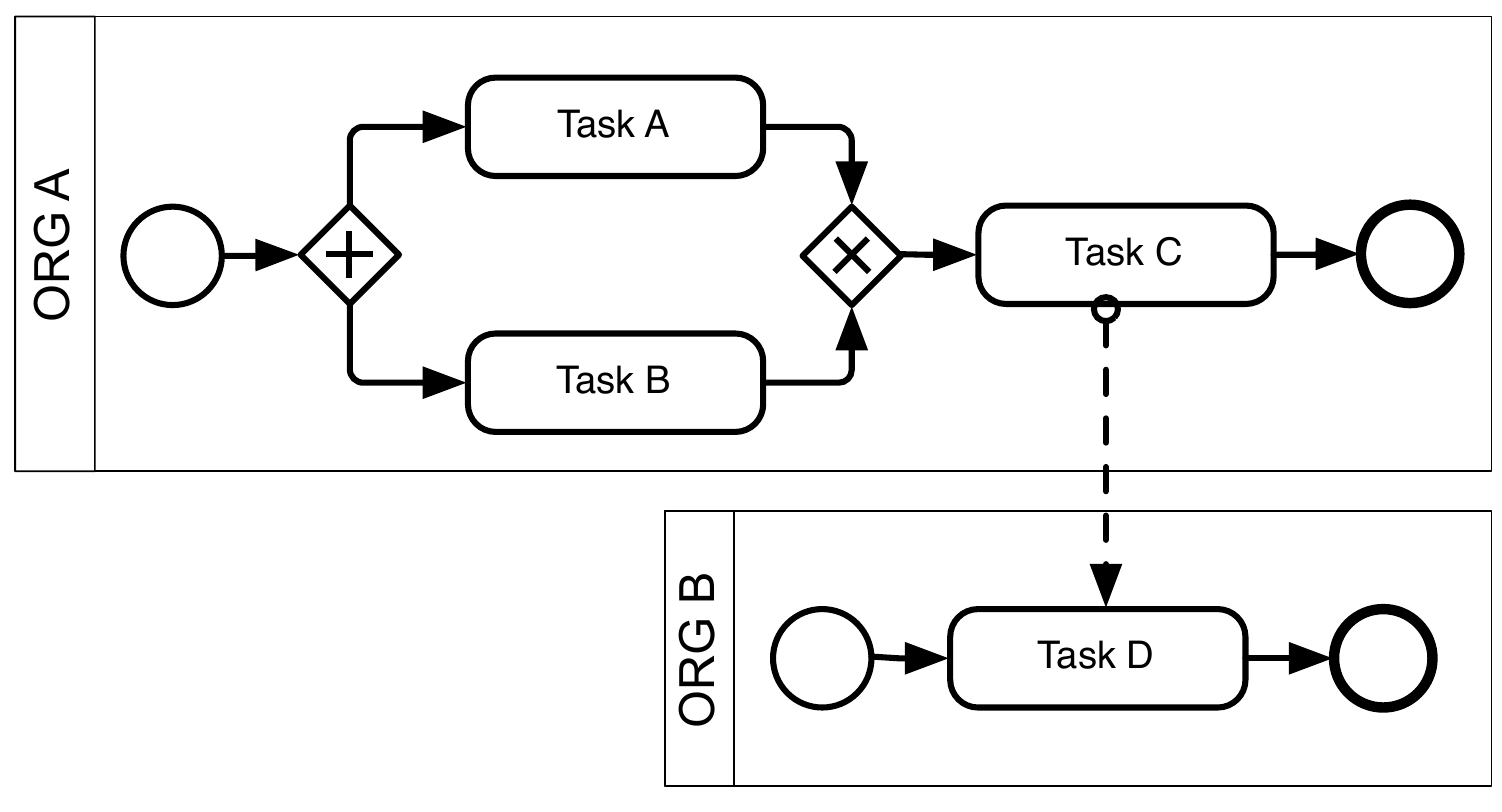}
    \vspace{-.3cm}
\caption{Unsound and message-relaxed unsound. \label{fig:classificationCollex1}}
\end{minipage}%
\hspace{-2mm}%
\begin{minipage}[c]{.33\textwidth}
%\vspace{-0.8cm}
\centering\setlength{\captionmargin}{0pt}%
  \includegraphics[width=0.9\textwidth]{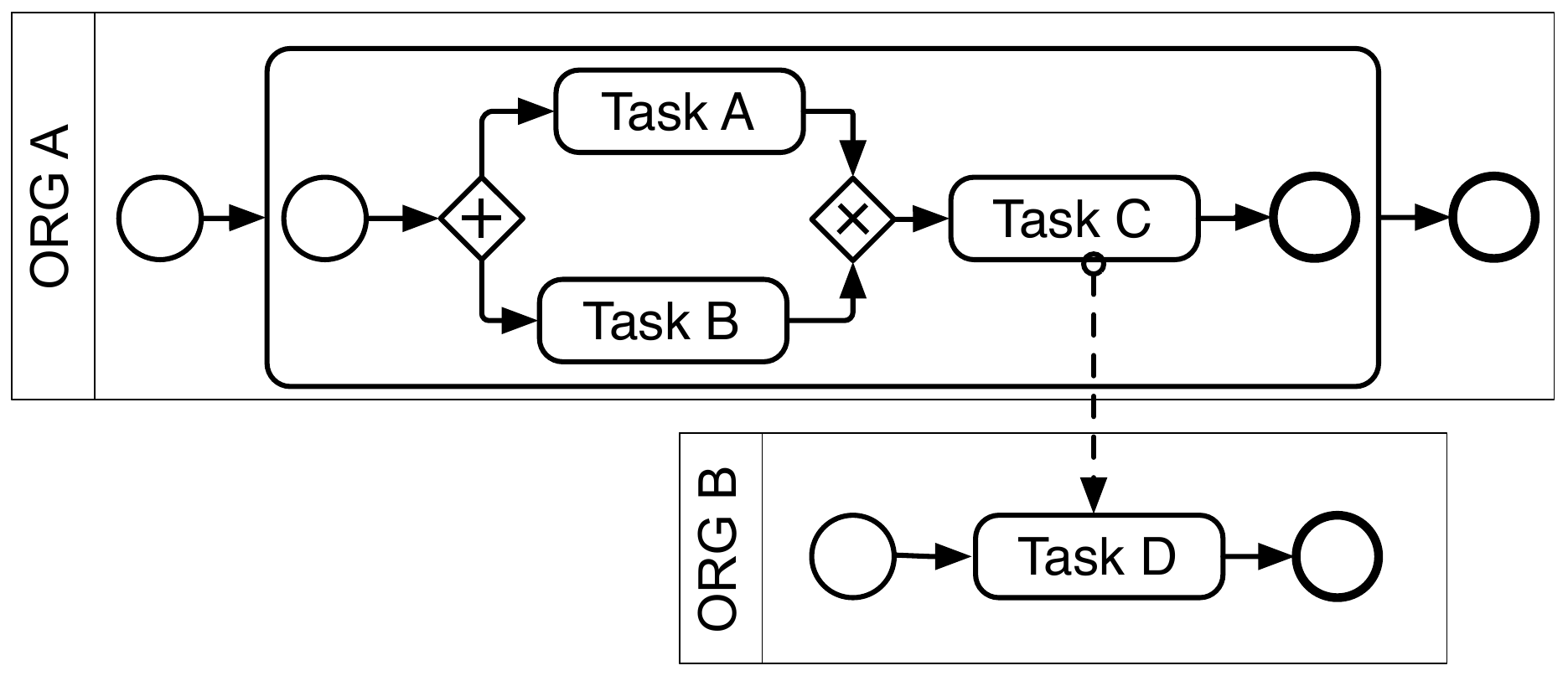}
    \vspace{-.3cm}
  \caption{Unsound and message-relaxed sound. \label{fig:classificationCollex2}}
\end{minipage}%
    \vspace{-.3cm}
\begin{minipage}[c]{.33\textwidth}
%\vspace{-1cm}
\centering\setlength{\captionmargin}{0pt}%
  \includegraphics[width=0.9\textwidth]{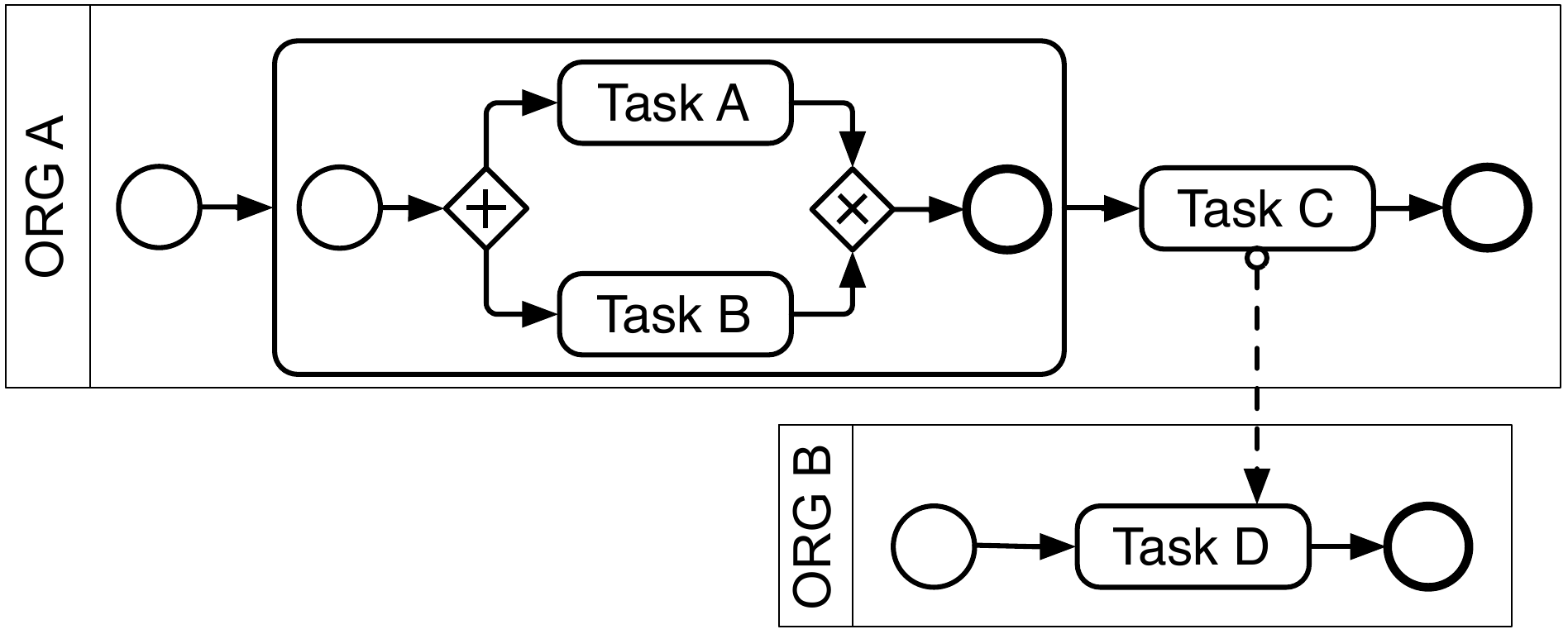}
    \vspace{-.3cm}
  \caption{Sound and message-relaxed sound.     \label{fig:classificationCollex3}}
\end{minipage}
   \vspace{-.1cm}
\end{figure}

Interesting situations also arise when focussing on the collaboration level,  as highlighted in  Fig.~\ref{fig:collsubproceffect}(b).
Worth to notice is the possibility to transform with a small change an unsound collaboration into a sound one. 
In Fig.~\ref{fig:classificationCollex1}, \ref{fig:classificationCollex2}  and~\ref{fig:classificationCollex3} we report simple examples showing  the impact of sub-processes. The models are rather similar, 
but according to our classification their behaviours are significantly different. 
The collaboration model in Fig.~\ref{fig:classificationCollex1} is both unsound and message-relaxed unsound. Now, let consider another model obtained from that in Fig.~\ref{fig:classificationCollex1} 
by introducing a sub-process. The resulting collaboration, shown in Fig.~\ref{fig:classificationCollex2}, is 
message-relaxed sound, since the use of the sub-process mitigates the problem, but it is unsound.
In fact, when the two processes reach a completed configuration, there will be a pending message, since task C sends two messages and only one will be consumed by task D. Notice also that the process in ORG~A is unsafe. Thus, even if sound, the whole collaboration is unsafe.
% However, apart for the token in the end event of ORG~A, no other pending sequence token needs to be processed.   This means that the collaboration is sound.  
Differently,  Fig.~\ref{fig:classificationCollex3} shows that enclosing within a sub-process the part of the model generating multiple tokens has a positive effect on  the soundness of the model. The collaboration is sound since only one message is sent and there are no pending messages. Message-relaxed soundness  is also satisfied since it is implied by soundness.
%

%% file: formal.tex
%!TEX root = ../Main_classification_Express.tex

\section{Formal Framework} 
\label{sec:formal}
This section presents our formal semantics of \bpmn{}  collaborations.
To simplify its formal definition, we rely on a textual representation 
of collaboration models. This permits to have more coincise and manageable semantic rules, 
making it easier to achieve the classification results.  The direct semantics proposed in this paper is inspired by \cite{CorradiniPRT15} \cite{CorradiniSCP}, but its technical definition is significantly different. In particular, configuration states are here defined according to a global perspective.

%\modif{Specifically, we first present the syntax and operational semantics we defined for a relevant subset of \bpmn{} elements. 
%%This subset includes the elements for describing sub-processes and message exchanges, which play a key role in our work, while are usually not considered or over-abstracted by other formalisations.
%In selecting the  elements we followed a pragmatic approach as, even if we deal with a restricted number of elements, we focus on those regularly used to design process models in practice (corresponding to less than 20\% of the \bpmn{} vocabulary \cite{muehlen_how_2008}).  Anyway, extending the framework to include further elements is not particularly challenging (even a tricky element as the OR-join can be conveniently added to our formalisation, see \cite{sofsem:2018}). }
%%\br{Andando a BPM siamo sicuri che possiamo cavarcela cosi con l'or-join?}
%%\chia{sar� il caso di rintrodurre lo startRcv e l'endSnd anche se non usato ma x omogeneit� visto che abbiamo messo l'event based?}
%
%%
%%%%%For the sake of presentation, some definitions, or part of them, are available in the online appendix [9].
  
%!TEX root = ../BPMN_BPClassification_paper.tex
\smallskip
\noindent
\textbf{Textual Representation of \bpmn{} Collaborations.}
Our textual representation of the structure of \bpmn{} collaboration models is defined by 
the BNF grammar in Fig.~\ref{fig:syntaxGlobal}. 
 In the proposed grammar, the non-terminal symbols $\collabs$ and $\procs$ %, and $\nodes$ 
 represent \emph{Collaboration} structures and \emph{Process} structures, respectively. The  two syntactic categories directly refer to the corresponding notions in \bpmn{}. %, the third one is an auxiliary notion introduced to simplify our studies to refer to process elements different from start, end, and terminate events. 
The terminal symbols, denoted by the ${\sf sans\ serif}$ font, are the typical elements of a \bpmn\ model, i.e.~pools, events, tasks, sub-processes and gateways. 
 %
%We are not proposing a new modeling formalism, but we are only using a textual notation for the \bpmn{} elements, which is more manageable than the graphical notation for writing operational rules. 
%
%It is worth noticing that our syntax is too permissive with respect to the \bpmn{} notation, as it allows to write collaborations that cannot be expressed in \bpmn{}. Limiting such expressive power would require to 
%extend the syntax (e.g., by imposing processes having at least one end event), thus complicating the definition of the formal semantics. However, this is not necessary in our work, as 
%we are not proposing an alternative modelling notation, but we are only using a textual representation 
%of \bpmn{} models, which is more manageable for writing operational rules than the graphical notation. 
%Therefore, in our analysis we will only consider terms of the syntax that are derived from \bpmn{}
%models.
%
Intuitively, a BPMN collaboration model is rendered in our syntax as a collection of pools and each pool contains a process. More formally, a collaboration $\collabs$ is a composition, by means of the $\collabspar$  operator, of pools of the form $\poolG{\orgname}{\procs}%{\msgList}%{\msgList_2} 
$ where: $\orgname$ is the name that uniquely identifies the pool, and  $\procs$ is the process included in the  pool.
%;  and $\msgList_1$ and $\msgList_2$ are the sets of incoming and outgoing message edges (ranged over by m), respectively. 

%\subsection{Syntax of \bpmn{} Collaborations}
%\label{subsec:globalcoll}
%
\begin{figure}[h]
\centering
\setlength{\belowcaptionskip}{-7pt}
%\vspace{-.0002cm}
\small
$
\begin{array}{|@{\ }rcl@{\ }|}
\hline
&&\\[-0.35cm]
\collabs  & ::=  & \poolG{\orgname}{\procs} \quad \mid\ \quad \collabs_1\!\!  \collabspar\! \collabs_2

  \\[.1cm]
\procs  & ::=  &  
\evStartG{\edgeG{\edgename_{enb}}}{\edgeG{\edgename_o}} 
\ \mid\  \evEndG{\edgeG{\edgename_i}}{\edgeG{\edgename_{cmp}}} 
\ \mid\  \evStartRcvG{\edgeG{\edgename_{enb}}}{\messagename}{\edgeG{\edgename_o}} 
\ \mid\  \evEndSndG{\edgeG{\edgename_i}}{\messagename}{\edgeG{\edgename_{cmp}}} 
\ \mid\ \evTerminateG{\edgeG{\edgename_i}} 
  \\[.1cm]
& \mid & 
 \eventbasedG{\edgeG{\edgename_i}}
 {({\messagename_1}, {\edgename_{o1}}),\ldots,({\messagename_h}, {\edgename_{oh}})} 
%\mid\    \nodes   \ \mid\    \procs_1 \!\!  \procspar\! \procs_2
\ \mid\ \andSplitG{\edgeG{\edgename_i}}{\edgeList_o} 
\ \mid\  \xorSplitG{\edgeG{\edgename_i}}{\edgeList_o}
\ \mid\  \andJoinG{\edgeList_i}{\edgeG{\edgename_o}}
%& \mid & 
%\modif{\orSplitG{\edgeG{\edgename_i}}{\edgeList_o}}
%\ \ \mid\ \  \modif{\orJoinG{\edgeList_i}{\edgeG{\edgename_o}}}
\\[.1cm]
% \ \mid\  \eventbasedG{\edgeG{\edgename_i}}{\msgList_1,\msgList_2} 

& \mid &  
 \xorJoinG{\edgeList_i}{\edgeG{\edgename_o}}
\ \mid\   \taskG{\edgeG{\edgename_i}}{\edgeG{\edgename_o}}
\ \mid\   \taskreceiveG{\edgeG{\edgename_i}}{\edgeG{\edgename_o}}{{\messagename}}
\ \mid\   \tasksendG{\edgeG{\edgename_i}}{\edgeG{\edgename_o}}{{\messagename}}
\ \mid\   \emptyG{\edgeG{\edgename_i}}{\edgeG{\edgename_o}}
\\[.1cm]
& \mid &  
 \evInterRcvG{\edgeG{\edgename_i}}{\edgeG{\edgename_o}}{{\messagename}} 
\ \mid\
 \evInterSndG{\edgeG{\edgename_i}}{\edgeG{\edgename_o}}{{\messagename}}
 \  \mid\  \subprocG{\edgeG{\edgename_i}}{\procs}{\edgeG{\edgename_o}} 
\ \mid\
 \procs_1 \!\!  \procspar\! \procs_2
 % \nodes_1\!\!  \nodespar\! \nodes_2
\\%[0.1cm]
%\msgList & ::=  &   \msg_1.\edgename_{1o}, \msg_2.\edgename_{2o} \ \ \mid\ \ \msg.\edgename_o, \msgList. \edgeList_o
%\msgList & ::=  &  ({\messagename}, {\edgename_o}) \ \ \mid\ \ \msgList_1,\msgList_2
%\\[.1cm]
\hline
\end{array}
$
\vspace{-.3cm}
\caption{BNF Syntax of \bpmn{} Collaboration Structures.}
\vspace{-.1cm}
\label{fig:syntaxGlobal}
\end{figure}

Notationally, $\messagename \in \msgSetM$ denotes a \emph{message edge}, enabling message exchanges between pairs of participants in the collaboration, %, while  $\msgList \in 2^\msgSetM$. %A set of messages can be also empty (e.g. considering a collaboration including only one pool). 
%Moreover, $\messagename$ denotes names uniquely identifying a message edge.
%
while $\edgename \in \edgeSetE$ denotes a \emph{sequence edge}, with $\edgeList \in 2^\edgeSetE$ denoting a set of sequence edges.
%; we require $ |\edgeList|>1$ when $\edgeList$ is used in joining and splitting gateways. 
%%
%Similarly, we require that an event-based gateway should contain at least two message events, i.e. 
%$h>1$ in each ${\sf eventBased}$ term.
%
For the convenience of the reader, we refer with $\edgename_i$ the edge incoming to an element, and with  $\edgename_o$ the edge outgoing from an element.
In the edge set $\edgeSetE$ we also include spurious edges denoting the enabled status of start events and the complete status of end events, named \emph{enabling} ($\edgename_{enb}$) and \emph{completing} ($\edgename_{cmp}$) edges, respectively.
They are needed to activate processes and sub-processes, as well as to check their completion. 
%
%
%
%The message exchange in a collaboration is modeled by means of \emph{message edges}. Here, they are represented by triples of the form $\msgEdge{\orgname_1}{\orgname_2}{\messagename}$ indicating, in order, the sending organization, the receiving organization and the message; we use $\msgSetMSG$ to denote the set of message edges.  
%
%
%  
%Moreover, as well as for the message edge we have that $\edgename$ denotes names uniquely identifying a sequence edge.
To achieve a compositional definition, each sequence (resp. message)  edge of the \bpmn{} model is split in two parts: the part outgoing from the source element and the part incoming into the target element. 
The two parts are correlated since edge names in the \bpmn{} model are unique.

The correspondence between the textual representation of \bpmn{} and the graphical one is straightforward. We exemplify it by means of a Travel Agency collaboration, used throughout the paper as a running example. 
%
% 
% \medskip
%\noindent
%\emph{Running Example (1/6).}\label{running}
%The  \bpmn{} collaboration model in Fig.~\ref{fig:runexample} represents a scenario where a Travel Agency continuously offers travels to a Customer, until an offer is accepted. If the Customer is interested in one offer, she books the travel and refuses other offers that the Travel Agency insistently proposes. As soon as the booking is received by the Travel Agency, it sends back a confirmation message, and asks for the payment. When this is completed, the ticket is sent to the Customer, and the Travel Agency activities immediately end.
%%
%\begin{figure}[t]
%  \centering
%    %\vspace{-.9cm}
%  \includegraphics[width=.85\textwidth]{img/RUNNING_new1}
%    \vspace{-.35cm}
%\caption{\bpmn\ collaboration diagram of a Travel Agency scenario.}
% \vspace{-.7cm}
%\label{fig:runexample}
%\end{figure}
%%
In particular, the textual representation of the scenario is as follows:

\begin{center}
$
\poolG{{\sf Customer}}{\procs_C}
%{\msgList_C}{\msgList_C'} 
\procspar \poolG{{\sf TravelAgency}}{\procs_{TA}}
%{\msgList_{TA}}{\msgList_{TA}'}
$
\vspace*{-.1cm}
\end{center}
with  
%$\msgList_C \!=\! {\msgList_{TA}} \!=\!({\sf Offer}, {\sf Travel}, {\sf Confirmation},  {\sf Payment}, {\sf Ticket})$ 
%
%$\msgList_C \!=\! {\msgList_{TA}'} \!=\!({\sf Offer}, {\sf Confirmation}, {\sf Ticket})$ and  $\msgList_C' \!=\! {\msgList_{TA}} \!=\! \!({\sf Travel}, {\sf Payment})$ 
%and 
$\procs_C$ defined as follows (process structure $\procs_{TA}$ is defined in a similar way):
$$
\begin{array}{l}
\\[-.7cm]
\evStartG{\edgeG{\edgename_0}}{\edgeG{\edgename_{1}}} 
%\evStart{\status_1}{\edge{\edgename_1}{\token_1}} 
\procspar 
 \xorJoinG{\{\edgename_{1},\edgename_{3}\}}{\edgeG{\edgename_{2}}}
%\xorJoin{\edge{\edgename_2}{\token_2}}{{\edge{\edgename_1}{\token_1}},{\edge{\edgename_3}{\token_3}}}  
\procspar
\taskreceiveG{\edgeG{\edgename_{2}}}{\edgeG{\edgename_{4}}}{{{\sf Offer}}}
%\taskRcv{\status_2}{\edge{\edgename_2}{\token_2}}{\edge{\edgename_4}{\token_4}}{\msgedge{{\sf Offer}}}{{\sf CheckTravelOffer}} 
\procspar
 \xorSplitG{\edgeG{\edgename_{4}}}{\{ \edgename_{3}, \edgename_{5}\}} 
 \\
%\xorSplit{\edge{\edgename_4}{\token_4}}{{\edge{\edgename_3}{\token_3}},{\edge{\edgename_5}{\token_5}}} 
\procspar
\tasksendG{\edgeG{\edgename_{5}}}{\edgeG{\edgename_{6}}}{{{\sf Travel}}}
%\taskSnd{\status_3}{\edge{\edgename_5}{\token_5}}{\edge{\edgename_6}{\token_6}}{{\sf Travel}}{{\sf BookTravel}}\\ 
\procspar
\evInterRcvG{\edgeG{\edgename_{6}}}{\edgeG{\edgename_{7}}}{{{\sf Confirmation}}} 
%\evInterRcv{\status_4}{\edge{\edgename_6}{\token_6}}{\edge{\edgename_7}{\token_7}}{\msgedge{\sf Confirmation}} \\
\procspar
\tasksendG{\edgeG{\edgename_{7}}}{\edgeG{\edgename_{8}}}{{{\sf Payment}}}\\
%\taskSnd{\status_5}{\edge{\edgename_7}{\token_7}}{\edge{\edgename_8}{\token_8}}{{\sf Payment}}{{\sf PayTravel}} 
\procspar
\evInterRcvG{\edgeG{\edgename_{8}}}{\edgeG{\edgename_{9}}}{{{\sf Ticket}}} 
%\evInterRcv{\status_6}{\edge{\edgename_8}{\token_8}}{\edge{\edgename_9}{\token_9}}{\msgedge{\sf Ticket}} \\
\procspar
\evEndG{\edgeG{\edgename_{9}}}{\edgeG{\edgename_{10}}} 
%\evEnd{\status_7}{\edge{\edgename_9}{\token_9}} 
\end{array}
\vspace*{-.3cm}
$$
%Process $\procs_{TA}$ is defined in a similar way. \\
%
%Moreover,  considering functions we define on the structure we can observe the following: 
%$\getParticipant{\poolG{{\sf Customer}}{\procs_{C}} 
%\procspar \poolG{{\sf TravelAgency}}{\procs_{TA}}}
%= \{\procs_{C}, \procs_{TA}\}$, 
%$ \getInitP{\procs_{C}} =  \{\edgename_0\} $, and 
%%$\getInitC{}$
%%$\getCmpP{}$
%$\getCmpC{\procs_{C}} = \{\edgename_{10}\}
%$. The others are defined in a similar way. 
%where, for simplicity, we identify edges $\edgename_i$ in progressive order (with $i= 0..10$).
%\qed

\smallskip
\noindent
\textbf{Semantics of \bpmn{} Collaborations.}
%\subsection{Semantics of \bpmn{} Collaborations}
%\label{subsec:globalcollsemantics}
%
The syntax presented so far permits to describe the mere structure of collaborations and processes. To describe its semantics we need to enrich it with a notion of execution state, defining the current marking of sequence and message edges. 
We use  \emph{collaboration configuration} and \emph{process configuration} to indicate these stateful descriptions.  
Formally, a collaboration configuration has the form 
$\langle \collabs, \stateedge, \statemsg \rangle$, where: 
$\collabs$ is a collaboration structure; 
$\stateedge$  is the part of the execution state at process level, storing for each sequence edge the current number of tokens marking it (notice it refers to the edges included in all processes of the collaboration); 
and $\statemsg$ is the  part of the execution state at collaboration level, 
storing for each message edge the current number of message tokens marking it. 
A process configuration, instead, has the form 
$\langle \procs, \stateedge \rangle$, where: 
$\procs$ is a process structure, and
$\stateedge$ is the execution state at process level. 
Specifically, a state $\stateedge : \edgeSetE \rightarrow \mathbb{N}$ is a function mapping edges to numbers of tokens ($\mathbb{N}$ is the set of natural numbers),
%The state obtained by updating in the state $\stateedge$ the number of tokens of the edge $\edgename$ to $\token$, written as 
%$\stateupd{\stateedge}{\edgename}{\token}$, is defined as follows: 
%$(\stateupd{\stateedge}{\edgename}{\token})(\edgename')$ returns $\token$ if $\edgename'=\edgename$, 
%otherwise it returns $\stateedge(\edgename')$. 
%The \emph{initial state}, where all edges are unmarked is denoted by $\initstateedge$ formally, $\initstateedge(\edgename)=0 \ \ \forall \edgename \in \edgeSetE$.
%
%
%
while a state  $\statemsg : \msgSetM \rightarrow \mathbb{N}$ is a function mapping message edges to numbers of message tokens, so that $\statefuncmsg{\messagename}=\token$ means that there are $\token$ messages of type $\messagename$ sent by a participant to another one that have not been received yet.
%
%Update for $\statemsg$ are defined in a way similar to  $\stateedge$'s definitions.
%
%\modif{Notably, to deal with decidability issues, in the implementation of the semantics the number of tokens marking  each sequence or message edge is bound.}
%
Predicate $\isInit{  \collabs, \stateedge, \statemsg  }$ holds if for each process $\procs$ included in $\collabs$ the predicate $\isInit{  \procs, \stateedge  }$ holds, and all message edges are unmarked in $\statemsg$.
Predicate $\isInit{  \procs, \stateedge  }$ holds when the start event of $\procs$ is enabled in $\stateedge$, i.e. it has a token in its enabling edge, while all other sequence edges (included the enabling edges of nested sub-processes) are unmarked in $\stateedge$.

Our operational semantics is defined by means of a \emph{Labelled Transition System}.  
Its transition relation on collaboration configurations formalises the execution of
message marking evolution according to the process evolution. 
The transition 
$\langle \collabs,\stateedge, \statemsg \rangle   \transitionColl{\poolLabel}{}  \langle \stateedge', \statemsg' \rangle$ means that  `the collaboration in the configuration $\langle \collabs, \stateedge, \statemsg \rangle$ can do a transition labelled by $\poolLabel$ and become the collaboration configuration $\langle \collabs, \stateedge', \statemsg' \rangle$ in doing so'. 
The definition of this relation relies on an auxiliary transition relation on process configurations, for which  
the transition $\langle \procs, \stateedge \rangle   \transitionColl{\procLabel}{}    \stateedge'  $
means that `the process in the configuration $\langle \procs, \stateedge \rangle$ can do a transition labelled by $\procLabel$ and become the process configuration $\langle \procs, \stateedge' \rangle$ in doing so'. 
We refer the interested reader 
to \cite{TR}  for a full account of the definition of these relations, while we report here 
by way of example a few operational rules: 
\vspace{-0,3cm}
$$
\begin{array}{c}
\begin{array}{r@{\ }c@{\ }l@{\qquad}l}
\langle \xorSplitG{\edgeG{\edgename}}{\{\edgeG{\edgename'}\} \cup\edgeList},\stateedge \rangle
& \transitionColl{\intAction}{} &
\incToken{\decToken{\stateedge}{\edgename}}{\edgename'}  
&
\textnormal{with}\ \statefunc{\edgename}>0
\\[.2cm]
\langle\andSplitG{\edgeG{\edgename}}{\edgeList}, \stateedge \rangle
            	&  \transitionColl{\intAction}{}  &  
  \incToken{\decToken{\stateedge}{\edgename}}{\edgeList}  
&
\textnormal{with}\    \statefunc{\edgename}>0
\\[.2cm]
\langle \taskreceiveG{\edgeG{\edgename}}{\edgeG{\edgename'}}{{\messagename}}, \stateedge \rangle
& \transitionColl{?\messagename}{} &
\incToken{ \decToken{\stateedge}{\edgename}}{\edgename'}
&
\textnormal{with}\ \statefunc{\edgename}>0
\end{array}
\\[.8cm]
\begin{array}{c@{\qquad\qquad}c}
   \prooftree 
   \langle \procs, \stateedge \rangle  \transitionColl{\internalAction}{}   \sigma'   
    \justifies 
    \begin{array}{r}
   \langle \poolG{\orgname}{ \procs}, \stateedge, \statemsg \rangle
   \transitionColl{\internalAction}{} 
\langle \stateedge',  \statemsg\rangle
             \end{array}
       \endprooftree   
&
      \prooftree 
   \langle \procs, \stateedge \rangle  \transitionColl{\receiveAction{\messagename}}{}   \sigma'    \quad \statefuncmsg{{\messagename}}>0   
    \justifies 
                \begin{array}{r}
   \langle \poolG{\orgname}{  \procs}, \stateedge, \statemsg \rangle
   \transitionColl{?\messagename}{} 
    \langle \stateedge',    \decToken{\statemsg}{\messagename}\rangle
             \end{array}
       \endprooftree   
\end{array}       
\vspace{-0.1cm}
\end{array}
$$
The first three rules concern the auxiliary transition relation on process configurations. 
In particular, the first one is applied when a token is available in the incoming sequence edge $\edgeG{\edgename}$ of a XOR split gateway.
The effect of the rule is to decrement in $\stateedge$ the number of tokens in the edge $\edgeG{\edgename}$ (by means of function $\mathit{dec}$) and increment (by means of function $\mathit{inc}$) the tokens of the edge $\edgeG{\edgename'}$ non-deterministically chosen from those outgoing from the gateway. 
Moreover,  a label $\intAction$, corresponding to a movement of tokens internal to the process, is observed.
The second rule is applied when there is at least one token in the incoming edge of an AND split gateway; as result of its application the rule decrements the number of tokens in the incoming edge and increments that in each outgoing edges.
The third rule acts similarly on the tokens marking the incoming and outgoing edges of a receiving task element, but it produces a label $?\messagename$ corresponding to the taking place of a receiving action. 
The other two rules, instead, refer to the transition relation on collaboration configurations. 
The first one simply states that when a process $\procs$ of a pool performs an internal action $\internalAction$ (i.e., an action $\intAction$ or a termination action $\morte$), the pool performs the corresponding internal action at collaboration layer.
Finally, given a pool including a process $\procs$, the last rule
can be applied only if $\procs$ is willing to perform a receiving action and 
there is at least one message $\messagename$ available. 
Of course, one message token is consumed by the transition (by means of function $\mathit{dec}$ applied to the state $\statemsg$). 
We will use $ {\transitionColl{}\!}^*$ to denote the reflexive and  transitive closure of the labelled transition relations introduced above.

%% file: wellstructure.tex
\section{Formalisation of BPMN Collaboration Properties}
\label{sec:Definitions}
We provide here a rigorous characterisation, based on the \bpmn{} formalisation presented so far, of the key properties studied in this work, both at the level of processes and collaborations.

\smallskip
\noindent
\textbf{Well-Structured \bpmn{} Collaborations.}
%\label{sec:WS}
Common process modelling notations, such as \bpmn{}, allow process models to have almost any topology. However, it is often desirable that models abide some structural rules. In this respect, a well-known property of a process model is that of \emph{well-structuredness}. 
In this paper we have been inspired by the definition of well-structuredness given by Kiepuszewski et al. \cite{kiepuszewskistructured2000}.  Such a definition was given on workflow models and it is not expressive enough for \bpmn{}, so we redefine it including all the elements defined in our semantics (i.e. not only base elements included in workflow models, but also the event-based gateway and sub-processes).
Technically, well-structuredness for processes is inductively defined on the process structure, 
by means of many cases (reported in \cite{TR}). Well-structuredness is then extended to collaborations, by requiring each process involved in a collaboration to be well-structured.

\begin{definition}[Well-structured collaborations] \label{WSC}
Let $\collabs$ be a collaboration, $\isWS{\collabs}$ is inductively defined as follows:
 $\isWS{\poolG{\orgname}{\procs}}$ if $\procs$ is well-structured;
and $\isWS{\collabs_1 \collabspar \collabs_2}$ if $\isWS{\collabs_1}$ and $\isWS{\collabs_2}$.
 \end{definition}

\noindent
\emph{Running Example.}\label{running}
Considering the proposed running example in Fig.~\ref{fig:runexample}, process $\procs_{C}$ is well-structured, while process $\procs_{TA}$ is not well-structured, due to the presence of the unstructured loop formed by the XOR join and an AND split. Thus, the overall collaboration is not well-structured.  
\qed

\smallskip
\noindent
\textbf{Safe \bpmn{} Collaborations.}
%
%\subsection{Safe \bpmn\ Collaborations}
%\label{sec:Safeness}
%
Another important condition usually required is \emph{safeness}. Its formal definition is based on the auxiliary function
$\maxMarking{\cdot}$ that, given a configuration $\langle \procs,  \stateedge \rangle$, 
determines the maximum number of tokens marking the sequence edges of 
elements in $\procs$ according to the state $\stateedge$.
We also need the following definition on the safeness of a process in a given state.
\begin{definition}[Current state safe process]\label{def:cs-safeP}
A process configuration $\langle \procs, \stateedge \rangle$ is current state safe (\emph{cs-safe}) if and only if $\maxMarking{  \procs, \stateedge  }\leq 1$.
\end{definition}
\noindent The definitions of safe processes and collaborations hence require that cs-safeness is preserved along the computations.  
\begin{definition}[Safe processes]\label{def:safenessP}
A process $\procs$ is \emph{safe} if and only if, given $ \stateedge$ such that $\isInit{  \procs, \stateedge  }$, 
for all $\stateedge'\!$  such that 
$ \langle \procs, \stateedge \rangle  {\transitionColl{}\!}^{*}   \stateedge' $ we have that $\langle \procs, \stateedge' \rangle$ is cs-safe.
\end{definition}

  \begin{definition}[Safe collaborations]\label{def:safenessC}
 A collaboration $\!\collabs$ is \emph{safe} if and only if, given  $ \stateedge$ and $\statemsg$  such that $\isInit{  \collabs, \stateedge, \statemsg  }$, 
for all  $\sigma'$ and $\delta'$ such that   
$\langle \collabs, \stateedge, \statemsg \rangle {\transitionColl{}\!}^{*}  \langle \stateedge', \statemsg' \rangle$  we have that $\maxMarking{  \collabs, \stateedge'  } \!\!\leq\!\!1$.
  \end{definition}

\noindent
\emph{Running Example.}\label{running}
Let us consider again our running example depicted in Fig.~\ref{fig:runexample}. Process $\procs_C$ is safe since there is not any process fragment capable of producing more than one token. Process $\procs_{TA}$ instead is not safe. In fact, if task {\sf Make Travel Offer} is executed more than once, we would have that the AND split gateway 
will produce more than one token in the sequence flow connected to the {\sf Booking Received} event. 
Thus, also the resulting collaboration is not safe. 
\qed%

\smallskip
\noindent
\textbf{Sound \bpmn{} Collaborations.}
%
%\subsection{Sound \bpmn\ Collaborations}
%\label{sec:soundness} 
%
Here we refer to the soundness as the need that from any reachable configuration it is possible to arrive in a (completed) configuration. This is possible under two different scenarios of a process in a given state. The first one is possible when all marked end events are marked exactly by a single token and all sequence edges are unmarked, while the second when no token are observed in the configuration (i.e. the case of termination due to a terminate event).  So, we  need the following definition of soundness of a process in a given state.
\begin{definition}[Current state sound process]\label{def:cssoundnessProc}
A process configuration  $\langle \procs, \stateedge \rangle$  is current state sound (\emph{cs-sound}) if and only if one of the following hold:
\begin{enumerate}[label=(\roman*)]
\item $\forall\, \edgename \in \marked{\stateedge\!}{\!\getCmpP{\procs}} \ . \ \stateedge(\edgename)=1$ and 
$\isZero{\procs, \stateedge}$;
\item 
$\forall\, \edgename \in \edges{\procs}  \ . \ \stateedge(\edgename)=0$.
\end{enumerate}
\end{definition}

\noindent In the definition above, we use the following auxiliary functions and predicates:
$\marked{\stateedge}{\getCmpP{\procs}}$  returns the set of edges of the end events with at least one token;
$\isZero{\procs, \stateedge}$ holds when all edges in $\procs$, except the completing ones, are unmarked;
%(hence, the case for the end event is $ \isZero{ \evEndG{\edgeG{\edgename_i}}{\edgeG{\edgename_{cmp}}},  \stateedge }$ if $\stateedge(\edgename_{i})=0$). 
$\edges{\procs}$ returns the edges in $\procs$.

\begin{definition}[Sound process]\label{def:soundnessProc}
A process $\procs$ is \emph{sound} if and only if, given $ \stateedge$ such that $\isInit{  \procs, \stateedge  }$, for all $\stateedge'\!$ such that 
$ \langle \procs, \stateedge \rangle  {\transitionColl{}\!}^{*} \stateedge' $ we have that there exists $\stateedge''\!$ such that  
$ \langle \procs, \stateedge' \rangle  {\transitionColl{}\!}^{*}  \stateedge'' $, 
and $ \langle \procs, \stateedge'' \rangle $  is cs-sound.
%
%$\forall\, \edgename_{cmp} \in \marked{\stateedge''\!}{\!\getCmpP{\procs}} \ . \ \stateedge''(\edgename_{cmp})=1$, and 
%$\isZero{\procs, \stateedge''}$
%$\oplus$
%$\forall\, \edgename \in \edges{\procs}  \ . \ \stateedge''(\edgename)=0$.
%
\end{definition}

\begin{definition}[Sound collaboration]\label{def:soundnessC} 
A collaboration $\collabs$ is \emph{sound} if and only if, given $\stateedge$ and $\statemsg$ such that $\isInit{  \collabs,  \stateedge, \statemsg  }$, for all  $\stateedge'$ and $\statemsg'$ such that 
$ \langle \collabs,  \stateedge, \statemsg \rangle  {\transitionColl{}\!}^{*} \langle \stateedge', \statemsg' \rangle$ we have that there exist $\stateedge''$  and $\statemsg''$ such that 
$ \langle \collabs, \stateedge', \statemsg' \rangle  {\transitionColl{}\!}^{*} \langle \stateedge'', \statemsg'' \rangle$, and $\forall\; \procs$ in $\collabs$ we have that  $ \langle \procs, \stateedge'' \rangle $  is cs-sound and $\forall\, \messagename \in \msgSetM\ .\  \statemsg''(\messagename) =0$.
%$\forall\, \edgename_{cmp} \in \marked{\stateedge''\!}{\!\getCmpP{\getParticipant{\collabs}}} \ . \ \stateedge''(\edgename_{cmp})=1$,  
%%$\isZero{\getParticipant{\collabs}, \stateedge''}$, 
%%
%$\forall\, \edgename \in \edges{\getParticipant{\collabs}\backslash \marked{\stateedge''\!}{\!\getCmpP{\getParticipant{\collabs}}} }  \ . \ \stateedge''(\edgename)=0$,
\end{definition}

Thanks to the capability of our formalisation of distinguishing sequence tokens from message tokens, we define the message-relaxed soundness property. It considers sound  those collaborations in which asynchronously sent messages are not properly handled by the receiver, so that at the completion of the execution there could be pending messages.
% need to be processed.
%
\begin{definition}[Message-relaxed sound collaboration]\label{def:RsoundnessC} 
A collaboration $\collabs$ is \emph{ Message-Relaxed  sound} if and only if, given $\stateedge$ and $\statemsg$ such that $\isInit{  \collabs,  \stateedge, \statemsg  }$, for all  $\stateedge'$ and $\statemsg'$ such that 
$ \langle \collabs,  \stateedge, \statemsg \rangle  {\transitionColl{}\!}^{*} \langle \stateedge', \statemsg' \rangle$ we have that there exist $\stateedge''$  and $\statemsg''$ such that 
$ \langle \collabs, \stateedge', \statemsg' \rangle  {\transitionColl{}\!}^{*} \langle \stateedge'',\statemsg'' \rangle$,
and $\forall\; \procs$ in $\collabs$ we have that  $ \langle \procs, \stateedge'' \rangle $  is cs-sound.
\end{definition}

\noindent
\emph{Running Example.}\label{running}
Let us consider again our running example. It is easily to see that process $\procs_C$  is sound, since it is always possible to reach the end event and when reached there is no token marking the sequence flows. Also process $\procs_{TA}$ is sound, since when a token reaches the terminate event, all the other tokens are removed from the edges by means of the killing effect. However, the resulting collaboration is not sound. In fact, when a travel offer is accepted by the customer, we would have that the AND-Split gateway will produce two tokens, one of which re-activates the task {\sf Make Travel Offer}. Thus, even if the process completes, the message lists of the customer may not be empty. However, the collaboration satisfies the message-relaxed soundness property. 
\qed

%% file: relationsonproerties.tex
%!TEX root = ../Main_classification_Express.tex

\section{Relationships among Properties}
\label{sec:rela}

In this section we study the relationships among the considered properties. In particular we investigate the relationship between (i) well-structuredness and safeness, (ii) well-structuredness and soundness, and (iii)  safeness and soundness. The proofs of these results are in the technical report~\cite{TR}.

\smallskip
\noindent
\textbf{Well-structuredness vs. Safeness in \bpmn.}
\label{subsec:safecompositionality}
%
%\subsection{Well-structuredness vs. Safeness in \bpmn}
%\label{sec:classification}
%
We present here some of the main results of this work concerning the correlation between well-structuredness and safeness, both at process and collaboration level. Specifically, we demonstrate that all well-structured models are safe  (Theorem~\ref{thmsafeness} and Theorem~\ref{collsafe}), and that the vice versa does not hold.
%
%To this aim, first we show that a process in the initial state is cs-safe (Lemma~\ref{init}).
%Then, we show that cs-safeness is preserved by the evolution of well-structured core process elements  (Lemma~\ref{nodesafe})
%and processes (Lemma~\ref{wssafe}). These latter two lemmas rely on the notion of \emph{reachable} 
%processes. \br{Va estesa la nozione di reacheability!} In fact, the syntax in Fig.~\ref{fig:syntaxGlobal} is too liberal, as it allows terms that cannot 
%be obtained (by means of transitions) from a process in its initial state. 
%%This last notion, in its own turn, needs the definition of initial state for a node.
%
The proof that well-structured processes are safe   is based on two auxiliary results: (a) a process in the initial state is cs-safe, and (b) cs-safeness is preserved by the evolution of well-structured processes.\\

\begin{theorem}\label{thmsafeness}
Let $\procs$ be a process, if $\procs$ is well-structured then $\procs$ is safe.
\end{theorem}
\vspace{-0.5cm}
\begin{proofS}
We show that if $\langle \procs,  \stateedge \rangle  \transitionColl{}{}^*\!\! \stateedge'$ then $\langle \procs,  \stateedge' \rangle$ is cs-safe,
by induction on the length $n$ of the sequence of transitions from $\langle \procs,  \stateedge \rangle$ to $	\langle \procs,  \stateedge' \rangle $. 
\qed
\end{proofS}\\

%We now extend the previous results to collaborations.
 
\begin{theorem}\label{collsafe}
%Let $\collabs$ be a collaboration, if $\isWS{\collabs}$  then $\collabs$ is safe.
Let $\collabs$ be a collaboration, if $\collabs$  is well-structured then $\collabs$ is safe.
\end{theorem}
\vspace{-0.5cm}
\begin{proofS}
We proceed by contradiction.  
\qed
\end{proofS}\\

\begin{figure}[t]
\centering
\begin{minipage}[c]{.50\textwidth}
%\vspace{-0.8cm}
\centering\setlength{\captionmargin}{0pt}%
  \includegraphics[width=0.6\textwidth]{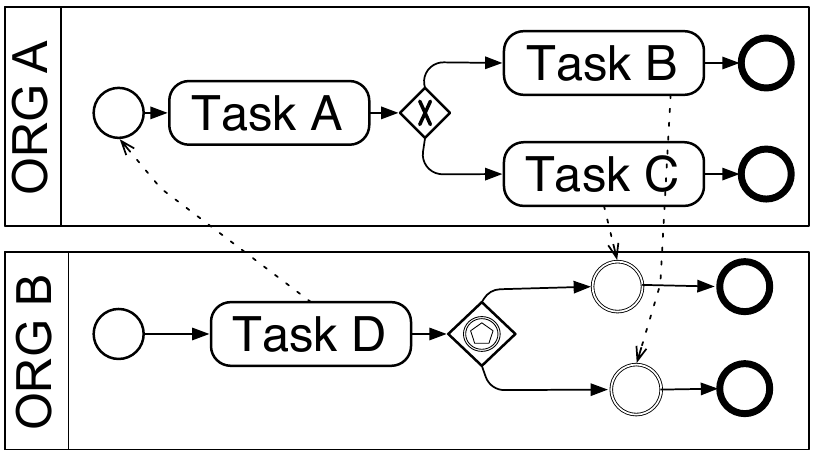}
    \vspace{-.3cm}
  \caption{A safe \bpmn\ collaboration not well-structured. \label{fig:example}}
\end{minipage}%
\hspace{-2mm}%
\begin{minipage}[c]{.50\textwidth}
%\vspace{-1cm}
\centering\setlength{\captionmargin}{0pt}%
\includegraphics[width=0.8\textwidth]{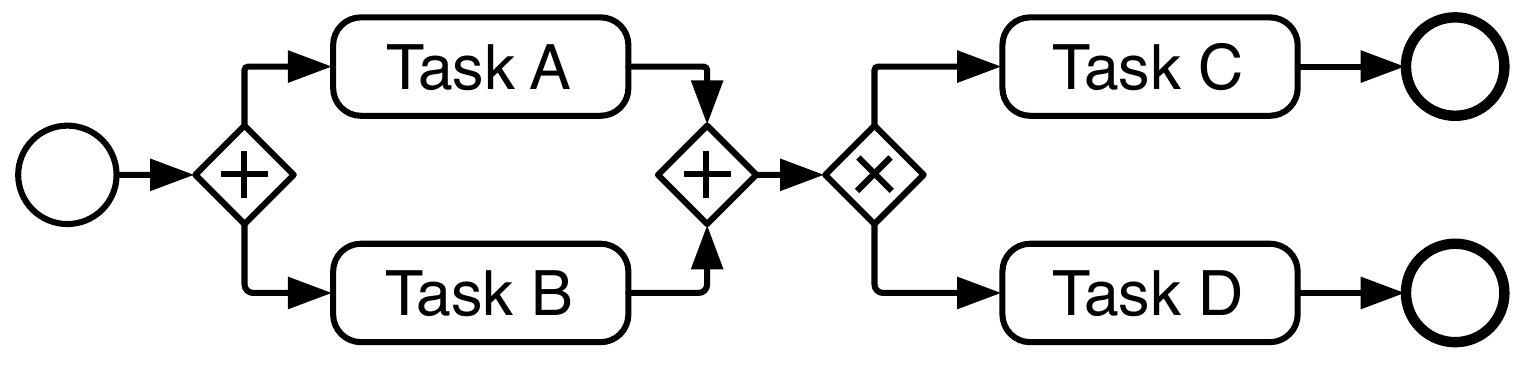}
    \vspace{-.3cm}
  \caption{An example of sound process not well-structured.     \label{fig:soundnows}}
\end{minipage}
   \vspace{-.1cm}
\end{figure}

The reverse implications of Theorems~\ref{thmsafeness} and~\ref{collsafe} are not true. 
In fact, there are safe processes/collaborations that are not well-structured.
The collaboration diagram represented in Fig.~\ref{fig:example} is an example. The involved processes 
are trivially safe, since there are not fragments capable of generating multiple tokens; however they are 
not well-structured.
%
%
%\begin{figure*}[htbp]
%\centering
%\includegraphics[width=0.4\textwidth]{img/soundnows}
%\vspace{-.4cm}
%\caption{An example of sound process not Well-Structured.}
%\label{fig:soundnows}
%\end{figure*}
%
%\begin{figure}[t]
%  \centering
%    %\vspace{-.9cm}
%  \includegraphics[width=0.4\textwidth]{img/ExampleTheorema1}
%    \vspace{-.25cm}
%\caption{A safe \bpmn\ collaboration not well-structured}
% \vspace{-.4cm}
%\label{fig:example}
%\end{figure}
%

\smallskip
\noindent
\textbf{Well-structuredness vs. Soundness in \bpmn.}
%
%\subsection{Well-structuredness vs. Soundness in \bpmn}
%\label{sec:wssound}
%
We 
%prove the relationship between well-structuredness and soundness. Specifically, we 
demonstrate that a well-structured process is always sound  (Theorem~\ref{thmwssound}).
%, but there are sound processes that are not well-structured. 
The proof of this result relies on the fact that well-structured processes can always complete their execution.

\begin{theorem}\label{thmwssound} 
Let $\procs$ be a well-structured process, then $\procs$ is sound.
\end{theorem}
\vspace{-0.5cm}
 \begin{proofS}
We proceed by case analysis.
\qed
\end{proofS}\\

The reverse implication of Theorem~\ref{thmwssound} is not true. In fact there are sound processes that are not well-structured; see for example the process represented in Fig.~\ref{fig:soundnows}. 
This process is surely unstructured, and it is also trivially sound, since it is always possible to reach an end event without leaving tokens marking its sequence flows.
%
%\begin{figure*}[htbp]
%\centering
%\includegraphics[width=0.4\textwidth]{img/soundnows}
%\vspace{-.4cm}
%\caption{An example of sound process not Well-Structured.}
%\label{fig:soundnows}
%\end{figure*}
%
What is particularly interesting is that Theorem~\ref{thmwssound} does not extend to the collaboration level. In fact, when we put well-structured processes together in a collaboration, this could be either sound or unsound. This is also valid for message-relaxed soundness.

\begin{theorem}\label{thmunsoundcoll}
Let $\collabs$ be a collaboration, $\isWS{\collabs}$ does not imply $\collabs$  is (message-relaxed) sound.
\end{theorem}
\vspace{-0.5cm}
\begin{proofS}
We proceed by contradiction.  
\qed
\end{proofS}\\
 
%
%
% 
%\begin{theorem}\label{thmunMDSsoundcoll}
%Let $\collabs$ be a collaboration, $\collabs$  is WS does not imply $\collabs$ is message-aware sound.
%\end{theorem}

\begin{figure}[b]
\centering
\begin{minipage}[c]{.50\textwidth}
%\vspace{-0.8cm}
\centering\setlength{\captionmargin}{0pt}%
  \includegraphics[width=0.7\textwidth]{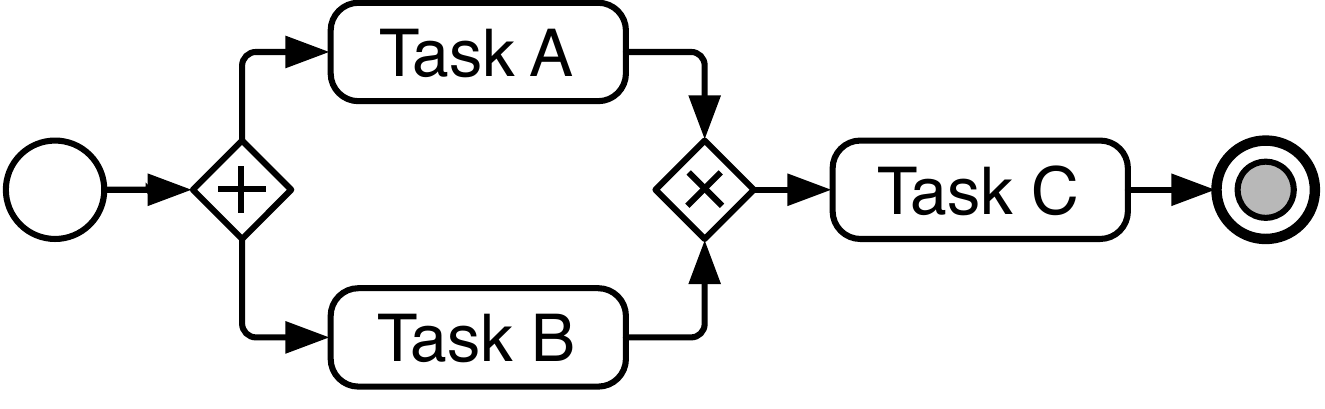}
    \vspace{-.3cm}
  \caption{An example of unsafe but sound process. \label{fig:terminate}}
\end{minipage}%
\hspace{-2mm}%
\begin{minipage}[c]{.50\textwidth}
%\vspace{-1cm}
\centering\setlength{\captionmargin}{0pt}%
  \includegraphics[width=0.7\textwidth]{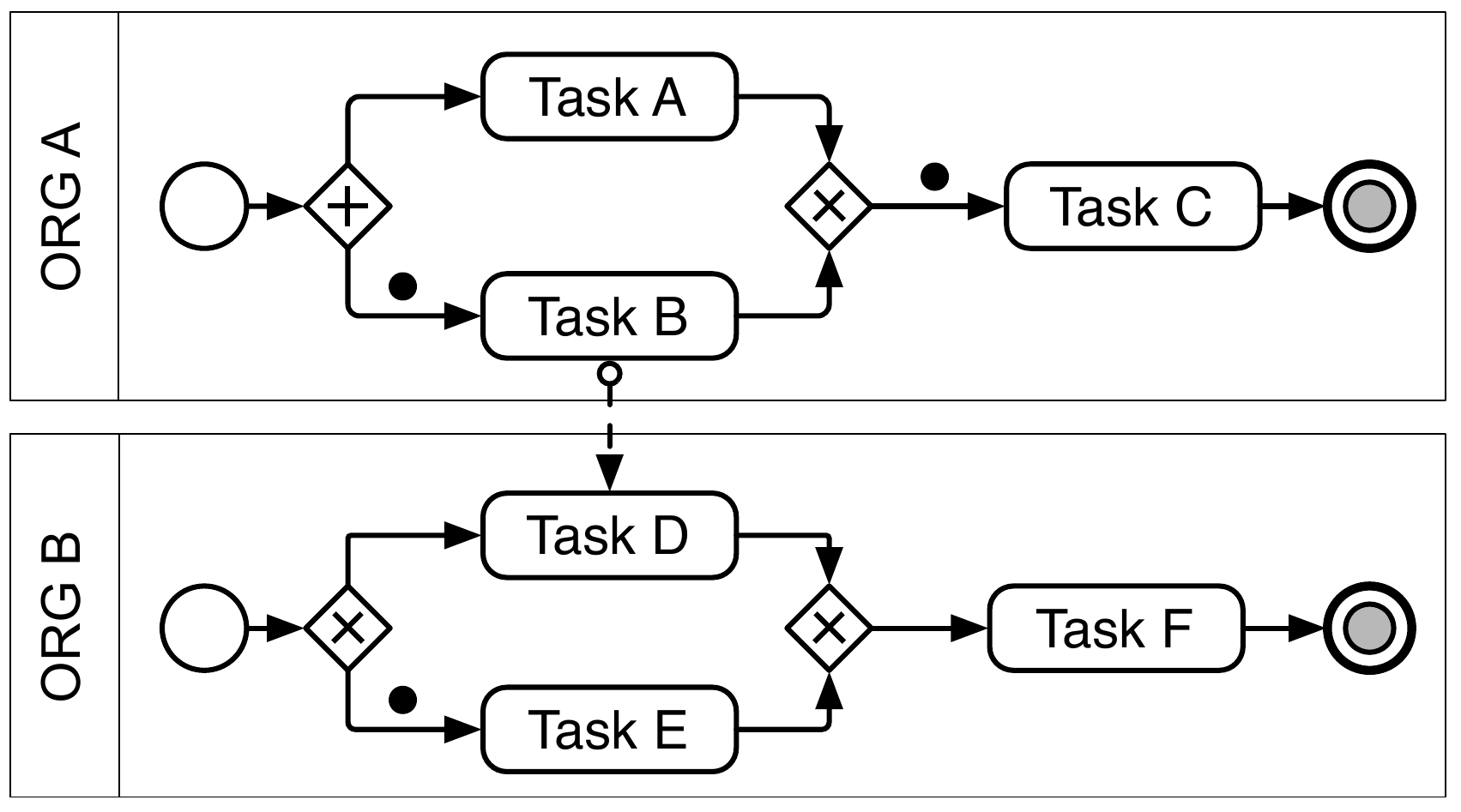}
    \vspace{-.3cm}
  \caption{An unsafe but sound collaboration.    \label{fig:killcoll}}
\end{minipage}
   \vspace{-.1cm}
\end{figure}

\smallskip
\noindent
\textbf{Safeness vs. Soundness in \bpmn{}.}
Finally, we present the relationship between safeness and soundness. Specifically we demonstrate that there are unsafe models that are sound. This is a peculiarity of \bpmn{}, faithfully implemented in our semantics, thanks to its capability of supporting the terminate event and (unsafe) sub-processes.
Theorem \ref{thmunsafesoundP} is demonstrated by the process in Fig. \ref{fig:terminate}, while  Theorem \ref{thmunsafesound} by the collaboration in Fig.   \ref{fig:killcoll}.

%
%
%\begin{figure*}[htbp]
%  \centering
%   % \vspace{-1.2cm}
%  \includegraphics[width=0.4\textwidth]{img/terminate}
%    \vspace{-.2cm}
%  \caption{An example of unsafe but sound process.}
%  \vspace{-.2cm}
%  \label{fig:terminate}
%\end{figure*}
%
% 
%\begin{figure*}[htbp]
%  \centering
%  \includegraphics[width=0.3\textwidth]{img/killcoll}
%    \vspace{-.2cm}
%  \caption{An unsafe but sound collaboration.}
%  \vspace{-.2cm}
%  \label{fig:killcoll}
%\end{figure*}

  \begin{theorem}\label{thmunsafesoundP}
Let $\procs$ be a process, $\procs$  is unsafe does not imply $\procs$  is unsound.
\end{theorem}
\vspace{-0.5cm}
\begin{proofS}
We proceed by contradiction.
\qed
\end{proofS}\\
 
 \begin{theorem}\label{thmunsafesound}
Let $\collabs$ be a collaboration, $\collabs$  is unsafe does not imply $\collabs$  is unsound.
\end{theorem}
\vspace{-0.5cm}
  \begin{proofS}
We proceed by contradiction.
\qed
\end{proofS}\\
 
 \medskip
\noindent
\emph{Running Example.}\label{running}
Considering the collaboration of our running example, the Customer process is both safe and sound, while the process of the Travel Agency is unsafe but sound, since the terminate event permits a successful termination of the process. The collaboration is not safe, and it is also unsound but message-relaxed sound, since there could be messages pending in the message lists.

%% file: compositionality.tex
%!TEX root = ../BPMN_BPClassification_TR_Restile.tex
\section{Compositionality of Safeness and Soundness}  
\label{sec:compostionality}

In this section we study  how the behaviour of processes affects that of the entire resulting collaboration. 
%In particular, we show the interrelationship between the studied properties at collaboration and at process level. 
%At process level we also consider the compositionality of sub-processes, investigating how sub-processes behaviour impacts on the safeness and soundness of process including them.  

\smallskip
\noindent
\textbf{On Compositionality of Safeness.}
\label{subsec:safecompositionality}
We show here that safeness is compositional, that is the composition of safe processes always results in a safe collaboration.

\begin{theorem}\label{comp}
Let $\collabs$ be a collaboration, if all processes in $\collabs$ are safe then $\collabs$ is safe.
\end{theorem}
\vspace{-0.5cm}
 \begin{proofS}
We proceed by contradiction. 
\qed 
\end{proofS}\\

The unsafeness  of a collaboration, instead, cannot be in general determined by information about the unsafeness of the processes that compose it. Indeed, putting together an unsafe process with a safe or unsafe one, the obtained collaboration could be either safe or unsafe.
%Let us consider now some cases.
 
%
 
\noindent
\emph{Running Example.}\label{running}
The collaboration in our example is composed by a safe process and an unsafe one.
In fact, focussing on the process of the Travel Agency, we can immediately see that it is not safe: the loop given by a XOR join and an AND split produces multiple tokens on one of the outgoing edges of the AND split. Now, if we consider this process together with the safe process of Customer, the resulting collaboration is not safe. Indeed, the XOR split gateway, which checks if the offer is interesting, forwards only one token on one of the two paths. As soon as a received offer is considered interesting, the Customer process proceeds and completes. 
Thus, due to the lack of safeness, the travel agency will continue to make offers to the customer, but no more offer messages arriving from the Travel Agency will be considered by the customer. 
%, so, we have problem of safeness and we also have a deadlock. Deadlock detection is out of scope of this paper, we remind the interested reader to \cite{onoda_definition_1999,awad_structural_2008}.
\qed

\smallskip
\noindent
\textbf{On Compositionality of Soundness.}
\label{subsec:soundcompositionality}
As well as for the safeness property, we show now that it is not feasible to detect the soundness of a collaboration by relying only on information about soundness of processes that compose it. However, the unsoundness of processes implies the unsoundness of the resulting collaboration.

\begin{theorem}\label{soundcomp}
Let $\collabs$ be a collaboration, if some processes in $\collabs$ are unsound then $\collabs$ is unsound.
\end{theorem}
\vspace{-0.8cm}
\begin{proofS} 
We proceed by contradiction. 
\qed
\end{proofS}\\

On the other hand, when we put together sound processes, the obtained collaboration could be either sound or unsound, since we have also to consider messages. It can happen that either a process waits for a message that will never be received or it receive more than the number of messages it is able to process.

%Let us consider some examples. \smallskip \\
%
\noindent
\emph{Running Example.}\label{running}
In our running example, the collaboration is composed by two sound processes.  In fact, the Customer process is well-structured, thus sound. Focussing on the process of the Travel Agency, it is also sound since when it completes the terminate event aborts all the running activities and removes all the tokens still present. % (more details will follow in Section \ref{sec:BPMNclassification}). 
However, the resulting collaboration is not sound, since the message lists could not be empty.
\qed

%% file: Related.tex
\section{Related Work}
\label{sec:Related}
In this paper we provide a formal characterisation of well-structuredness for \bpmn{} models. 
To do that we have been inspired by the definition of well-structuredness given in \cite{kiepuszewskistructured2000}.  Other attempts are also available in the literature \cite{vanderaalstapplication1998,el-sabercmmi-cm2015},  but none of them has been extended to the collaboration level. 
%
%Van der Aalst et al. \cite{vanderaalstapplication1998} state that a workflow net is well-structured if the split/join constructions are properly nested. 
%
%El-Saber and Boronat \cite{el-sabercmmi-cm2015}  propose a formal definition of well-structured processes, in terms of a rewriting logic, but they do not extend this definition at collaboration level. 
%

We also consider the safeness property. Dijkman et al. \cite{dijkmansemantics2008} discuss about safeness in Petri Nets resulting from the translation of \bpmn{} models. In that work, safeness means that no activity will ever be enabled or running more than once concurrently. This definition is given using natural language, while in our work we give a precise characterisation of safeness for both processes and collaborations. Other approaches introducing mappings from BPMN to formal languages, such as YAWL \cite{deckertransforming2008}, %and COWS \cite{prandiformal2008}, 
do not consider safeness. %, even if its importance is underlined.

Moreover, soundness is considered as one of the most important correctness criteria. 
There is a plethora of different notions of soundness in the literature, referring to different process languages and even for the same process language, e.g. for EPC a soundness definition is given by Mendling in \cite{mendlingdetection2007}, and for Workflow Nets by van der Aalst  \cite{vanderaalstsoundness2011} provides two equivalent soundness definitions.
However, these definitions cannot be directly applied to \bpmn.
In fact, although the \bpmn\ process flow resembles to some extent the behaviour of Petri Nets, they are not the same. \bpmn\ 2.0 provides a comprehensive set of elements that go far beyond the definition of mere place/transition flows and enable modelling at a higher level of abstraction. 
For example, using Petri Nets it is difficult to describe certain operations typical of the business process domain, such as the termination event, and often it is required to rely on some limiting assumptions (e.g., safeness and well-structureness).

Other studies that try to characterise inter-organisational soundness are available \cite{vanderaalstprocess-oriented1999,roaverification2011}.
%A first attempt was done using a framework based on Petri Nets \cite{vanderaalstprocess-oriented1999}. The authors investigate IO-soundness presenting an analysis technique to verify the correctness of an interorganizational workflow. However, the study is restricted to structured models. 
%%
%Soundness regarding collaborative processes is also given in \cite{roaverification2011} in the field of the Global Interaction Nets, in order to detect errors in technology-independent collaborative
%business processes models.
However, differently from our work, these approaches do not apply to \bpmn{}, which is the modelling notation aimed by our study. 
%that we aim at study in our work we focus on \bpmn\  aiming at providing a classification of \bpmn\ collaborations according to the properties they satisfy.
In particular, our investigation of properties at collaboration level provides novel insights with respect to the state-of-the-art of \bpmn\ formal studies.

%% file: conclusions.tex
%!TEX root = ../BPMN_safeness_TR.tex
\section{Concluding Remarks}
\label{sec:conclusion}

Our study formally defines some important correctness properties, namely well-structuredness, safeness, and soundness, both at the process and collaboration level of \bpmn{} models.  
We demonstrate the relationships between the studied properties, with the aim of classifying \bpmn\ collaborations according to the properties they satisfy. Rather than converting the BPMN model to a Petri or a Workflow Net and studying relevant properties on the model resulting from the mapping, we directly define  such properties on \bpmn, thus dealing with its complexity and specificities directly. Our approach is based on a uniform formal framework and is not limited to models with a specific topology, i.e., models do not need to be block-structured.